\documentclass[aps,prl,superscriptaddress,twocolumn,bibliography]{revtex4-1}
\usepackage{amsfonts}
\usepackage{amstext}
\usepackage{amssymb}
\usepackage{amsmath}
\usepackage[utf8]{inputenc}
\usepackage{graphicx}
\usepackage{graphics}
\usepackage{cancel}
\usepackage{color,xcolor}

\usepackage{hyperref}
\hypersetup{colorlinks=true,citecolor=blue,linkcolor=blue,filecolor=blue,urlcolor=blue,pdfstartview=FitH,pdfpagemode=UseNone}
\usepackage{tikz}

\usepackage{stix}
\definecolor{boxcolor}{HTML}{108f64}

\bibpunct{[}{]}{,}{n}{}{}

\begin{document}

\title{Structure-driven phase transitions in paracrystalline topological insulators}

\author{Victor Regis}
\affiliation{Jožef Stefan Institute, 1000 Ljubljana, Slovenia}

\author{Victor Velasco}
\affiliation{Instituto de F\'isica, Universidade Federal do Rio de Janeiro, Caixa Postal 68528, 21941-972 Rio de Janeiro, Brazil}
\affiliation{School of Pharmacy, Physics Unit, Università di Camerino, Via Madonna delle Carceri 9, 62032 Camerino, Italy}

\author{Marcello B. Silva Neto}
\affiliation{Instituto de F\'isica, Universidade Federal do Rio de Janeiro, Caixa Postal 68528, 21941-972 Rio de Janeiro, Brazil}
\affiliation{Laboratório Nacional de Nanotecnologia, CNPEM, 13083-100, Campinas, São Paulo, Brazil}

\author{Caio Lewenkopf}
\affiliation{Instituto de F\'isica, Universidade Federal Fluminense, 24210-346 Niter\'oi, Brazil}

\date{\today}

\begin{abstract}
We study phase transitions driven by structural disorder in noncrystalline topological insulators. We introduce a procedural generation algorithm, the Perlin noise, typically used in computer graphics, to incorporate disorder to a two-dimensional lattice, allowing a continuous interpolation between a pristine and a random gas system, going through all different intermediate structural regimes, such as the paracrystalline and the amorphous phases. We define a two-band model, including intraorbital and interorbital mixings, on the structures generated by the algorithm and we find a sequence of structure-driven topological phase transitions characterized by changes in the topological Bott index, at which the insulating gap dynamically closes while evolving from the Bragg planes of the Brillouin zone towards the center. We interpret our results within the framework of Hosemann's paracrystal theory, in which distortion is included in the lattice structure factor and renormalizes the band-splitting parameter. Based on these results, we ultimately demonstrate the phenomenon of topological protection at its extreme.

\end{abstract}

\maketitle

%%%%%%%%%%%%%%%%%%%%%%%%%%%%%%%%%%%%%%%%%%%%%%%%%%
{\it Introduction.--} 
%%%%%%%%%%%%%%%%%%%%%%%%%%%%%%%%%%%%%%%%%%%%%%%%%%
Topological properties of quantum matter are a subject of intense experimental and theoretical research, both from the point of view of fundamental science and technological applications 
% through the synthesis of advanced materials 
\cite{Hasan2010,Qi2011}. Topological insulators (TIs) are materials characterized as being bulk insulators while sustaining metallic states on their surfaces, the so-called “bulk-edge” correspondence \cite{TKNN, Hatsugai1990-1, bernevig}. 
Unlike in ordinary insulators, these edge states are protected against disorder by topology and symmetry. 
Owing to this unusual feature, the electronic and spin transport in TIs exhibit very peculiar characteristics \cite{kane-mele, kane-mele-z2, bhz, Konig2007, Roy2009, bernevig}. 
%
% Since this class of materials does not fit the standard classification of solids into metals, 
% semiconductors or insulators, TIs open a new chapter in condensed matter research.

Translation symmetry is an essential element in formulating the theory of TIs.
For instance, the characterization of TI phases is based on the calculation of topological invariants \cite{Kohmoto1985, Hatsugai1990-1, Qi2008, Moessner2021}, which 
%do not depend on the local physics, such as the distribution of impurities, and 
involve the integration of certain quantities over the first Brillouin zone (BZ). Furthermore, distinct topologically protected states cannot be smoothly deformed into one another without a phase transition, if the deformation preserves symmetry \cite{bernevig}. 
The natural question that arises is if 
% whether or not 
strongly disordered and/or non-crystalline systems can exhibit topological properties. 
Studies on strong scalar disorder \cite{Li2009, Groth2009} and more recently on random lattices \cite{Agarwala2017,Mitchell2018,Varjas2019, Yang2019}, quasi-crystals \cite{Huang2018a, Huang2018b}, amorphous 
% to-be 
structures \cite{Marsal2020,Agarwala2020} support a positive answer to this question. 
Of special interest are amorphous TIs \cite{Corbae2023b}, due to the possibility of material realizations \cite{Costa2019,Focassio2021a,Corbae2023a}.
However, the lack of studies on the electronic and topological properties as a function of the amorphization process cast shadows 
% to what is really going on for some of 
on the nature of 
its intermediate phases, such as the paracrystalline \cite{hosemann} or glassy quantum matter \cite{Sahlberg2020},
% These works also raise interesting new issues concerning the existence, or not, of real materials that display topological properties in the amorphous phase, or
and keep open the issue on 
how to characterize a topological phase transition in the amorphization process of a TI. 

% The simulation of noncrystalline lattices, for instance amorphous and fully disordered structures, 
% is usually performed through 
Noncrystalline lattices are usually simulated using 
pseudo-random sampling or the Voronoi tessellation method \cite{Florescu2009, Klatt2018, Park2007, Ruocco1991, Torres2010, Derlet2020, Ingo2021}. 
However, neither of these methods allow for a smooth transition from high- to low-correlations between neighboring sites.
% since it generates static amorphous systems. 
Alternatively, the bond-flipping technique \cite{Costa2019,Focassio2021a} and molecular dynamics \cite{Focassio2021b} account for the relaxation of the crystalline structure, which eventually leads to an amorphous lattice, but since it relies on {\it ab initio} calculations, its computational cost does not allow a systematic study of intermediary stages along the process. 
%This points to the need of introducing new techniques of gradual transitions between pristine and 
% random systems in the study of structural disorder in TIs. 
Hence, new techniques for study of a gradual increase of the structural disorder in TIs are on demand.

We develop an approach to overcome this difficulty based on a computer graphics procedure. 
In this field, the goal is to accurately produce computer-generated textures and/or real world landscapes with optimized efficiency, that is achieved by {\it Procedural Generation} algorithms, such as L-systems, Perlin Noise and diamond-square algorithms \cite{procedural-book}. 
Here, we use the Perlin Noise (PN) procedure \cite{Perlin1985, improved-perlin}, 
% designed to create procedural textures for computer-generated effects, 
which unlike simple random noise, produces smooth and continuous 
%, and self-similar 
outputs. 
Moreover, the PN algorithm creates correlated noise maps, which for the purposes of establishing structural disorder in a lattice, guarantees a smooth evolution of the atomic positions.

In this Letter, we introduce the use of the PN to realize the crossover from pristine lattices to random systems in two-dimensions (2D) by continuously increasing the structural disorder. 
% This allows us cover a plethora of intermediate structural phases in the paracrystalline regime \cite{hosemann}. 
By considering a two-band model 
% in the lattices generated by the PN, 
we characterize a series of structure-driven topological phase transitions by using the Bott index ${\cal B}$ as a topological marker \cite{Huang2018a, Huang2018b, Loring_2010}. 
% Additionally, 
We interpret our results in terms of Hosemann's paracrystal theory \cite{hosemann}, which enables us to introduce the effects of structural disorder in the lattice structure factor that perturbatively renormalize the parameters, and to ultimately demonstrate topological protection at its extreme in the amorphization process.

%%%%%%%%%%%%%%%%%%%%%%%%%%%%%%%%%%%%%%%%%%%%%%%%%%
{\it Perlin noise.--}
%%%%%%%%%%%%%%%%%%%%%%%%%%%%%%%%%%%%%%%%%%%%%%%%%%
The PN algorithm can be applied in iteration steps and is divided in two parts: randomization and amorphization \cite{SupMat}. The randomization process is the generation of the noise map following the PN procedure. One starts by dividing the 2D space into $N^2$ square cells of side $a$, then one inserts the coordinates $(x, y)$ of the points into a square lattice configuration. The noise $n(x_i,y_i, t)$ of each coordinate point at iteration $t$ is then generated by the PN algorithm: it identifies the cell containing the point and returns a noise map at each $t$. 
%For the sake of simplicity, we used as input coordinates the positions of the points in a square lattice for every iteration, but as this is a pseudo-random procedure, we ended up generating $t$ different noise maps, one for each step.

%%%%%%%%%%%%%%%%%%%%%%%%%%%%%%%%%%%%%%%%%%%%%%%%%%%%%%%%%%%%%%%%%%%%%%%%
\begin{figure}[!t]
\includegraphics[width=0.98\columnwidth]{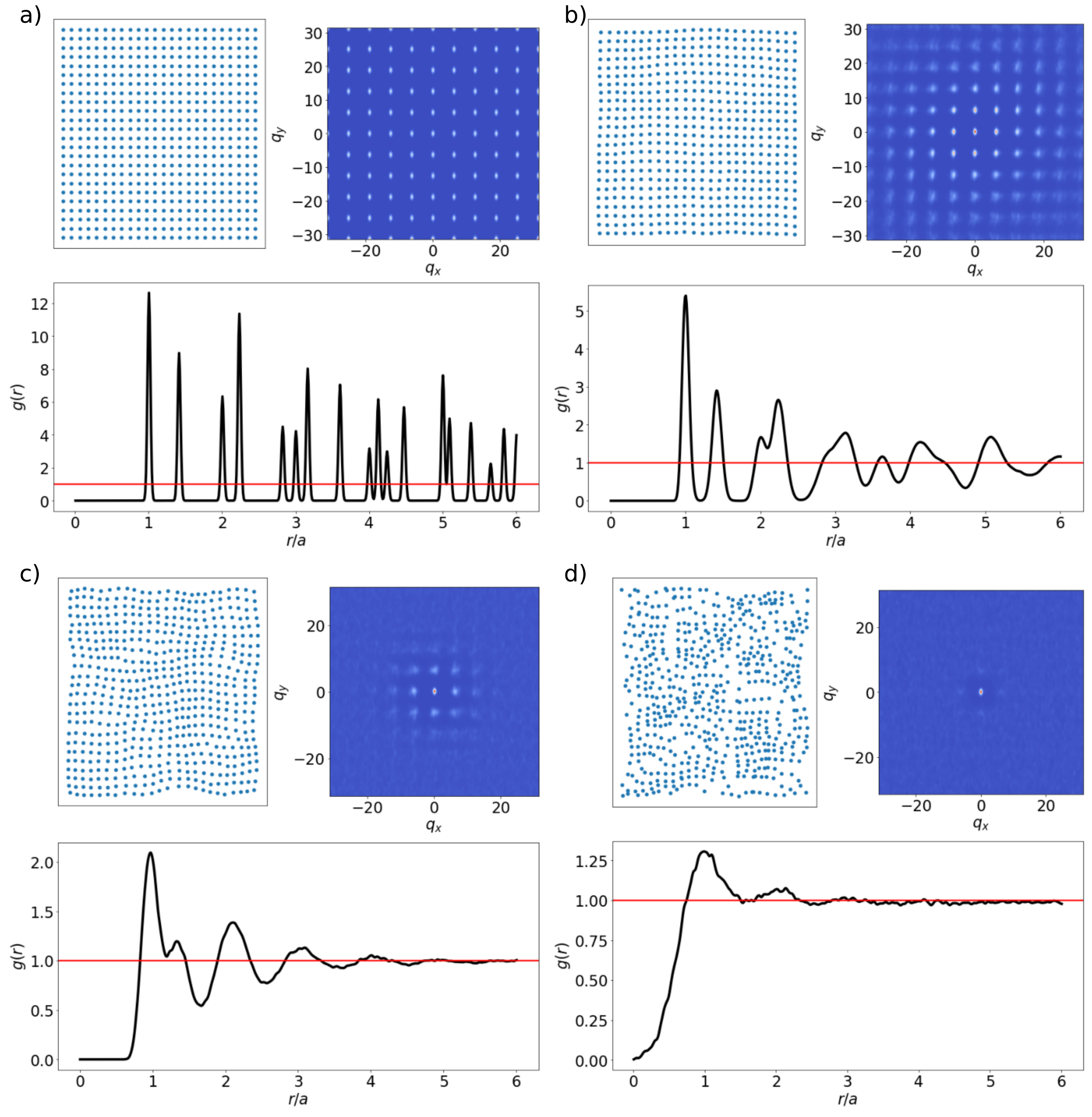}
\caption{Crystalline-random transition in a $24\times 24$ 2D system using the PN algorithm. 
(a) Square lattice $(t = 0)$, (b) paracrystal $(t \approx 30)$, (c) amorphous $(t \approx 70)$, and 
(d) random lattice $(t= 100)$. 
Each panel contains a representative lattice realization, the average 
% structure factor 
$S(\mathbf{q})$, and the average 
% radial distribution function 
$g(r)$, the averages taken over $\sim 10^2$ realizations.}
\label{Fig:Structures}
\end{figure}
%%%%%%%%%%%%%%%%%%%%%%%%%%%%%%%%%%%%%%%%%%%%%%%%%%%%%%%%%%%%%%%%%%%%%%%%

The amorphization process is responsible for 
%dynamically 
changing the positions of the lattice sites following the generated noise map at each iteration. 
From $n(x,y,t)$, an offset value $\delta$ is drawn, which is added to each position $(x_i, y_i)$ of the points in the system. 
These trial displacements can be accepted or not, depending on whether there is not a neighbor site within a distance set by the lattice parameter. 
Even if the criteria is not satisfied, we add a small offset to the position in order to guarantee that in the end of the $t$ iterations steps, long-range correlation is completely absent in the system.
%, retaining only short-range correlations. 
Finally, to ensure the complete transition from a pristine crystal to a random configuration, we let the sites to be offset independently of their neighbors and linearly increase the offset weight in the last half of the amorphization process \cite{SupMat}.

Figure~\ref{Fig:Structures} illustrates four structural phases obtained by applying the PN algorithm up to $t = 100$ steps.
The systems are characterized by their structure factors $S(\mathbf{q})$ and radial distribution functions (RDF) $g(r)$. 
The crystalline phase in Fig.~\ref{Fig:Structures}(a) exhibits a textbook crystal diffraction pattern, with pronounced $\delta$-like peaks both in $S(\mathbf{q})$ as in $g(r)$. 
The paracrystalline phase, Fig. \ref{Fig:Structures}(b), in turn, is characterized by the progressive loss of long-range correlation in $S(\mathbf{q})$, consistent with the broadening of the diffraction peaks with larger $r/a$ in the RDF. 
In the amorphous phase, Fig.~\ref{Fig:Structures}(c), there is no significant symmetry left in the system, as clearly inferred from the RDF, that displays one large peak corresponding to nearest-neighbors sites and a couple of broader peaks that quickly wane with increasing $r$. 
Lastly, for the random lattice, Fig.~\ref{Fig:Structures}(d), the system completely loses correlations, characterized by a single, trivial peak at $S({\bf 0})$ and by $g(r > a)$ quickly converging to unity.

%%%%%%%%%%%%%%%%%%%%%%%%%%%%%%%%%%%%%%%%%%%%%%%%%%%%%%%%%%%
{\it The model.--} 
%%%%%%%%%%%%%%%%%%%%%%%%%%%%%%%%%%%%%%%%%%%%%%%%%%%%%%%%%%%
Let us consider the $2D$ Agarwala-Shenoy model (2DAS) \cite{Agarwala2017}, 
a two-orbital tight-binding model, on lattices generated by the PN algorithm. 
This model is an extension of the half-BHZ model \cite{bhz}, 
originally designed for the band-inversion of $s,p$ orbitals in HgTe/CdTe heterostructures, 
that
% the 2DAS model 
also includes an interorbital mixing $\lambda$ and an intraorbital hopping $t_2$, 
% in such a way that it possesses no internal symmetries like time reversal, charge conjugation or sublattice. 
that can break time-reversal, charge conjugation, and/or sublattice symmetries \cite{Agarwala2017, SupMat}.
The Hamiltonian is 
\begin{equation}
{\cal H}=\sum_{I,J}\sum_{\alpha,\beta}t_{\alpha\beta}({\bf r}_{IJ})c^\dagger_{I,\alpha} c^{}_{J,\beta},
\label{eq:Hamiltonian}
\end{equation}
where $c^\dagger_{I,\alpha} (c_{J,\beta})$ are creation (annihilation) operators of electrons on sites $I, J$ with spin (orbital) indices $\alpha, \beta$. ${\bf r}_{IJ}={\bf r}_I-{\bf r}_J$ is the distance vector between sites on the lattice, and 
% the generalized hopping matrix 
$t_{\alpha\beta}({\bf r}_{IJ})$ contains an on-site energy 
\begin{equation}
    t_{\alpha\beta}({\bf 0})=\epsilon_{\alpha\beta}=
    \begin{pmatrix}
        2+M & (1-i)\lambda \\
        (1+i)\lambda &  -(2+M)
    \end{pmatrix},
\label{eq:agarwala-H}
\end{equation}
and a hopping matrix, $t_{\alpha\beta}({\bf r}_{IJ})=t(|{\bf r}_{IJ}|)\,T_{\alpha\beta}(\hat{\bf r}_{IJ})$. 
Here, $M$ is the band-splitting parameter associated with the energy difference between the $s$ and $p$ orbitals. 
The radial part, $t(r) = C\Theta (R-r)e^{-r/a}$, where $\Theta(r)$ ensures that hoppings are only allowed for $r<R$, $C$ is a constant, and $a$ is the square lattice spacing. 
Finally, $T_{\alpha\beta}(\hat{\bf r})$ reads
%
%\begin{widetext}
\begin{equation}
    T_{\alpha\beta}(\hat{\bf r})=\frac{1}{2}
    \begin{pmatrix}
        -1+t_2 & -i\, e^{-i\theta}+ \alpha(\theta) \\
        -i\, e^{i\theta}+ \alpha^*(\theta) & 1+t_2
    \end{pmatrix},
\label{eq:agarwala-hopping}
\end{equation}
where $\alpha(\theta)= \lambda \sin^2\theta(1+i)-1$ and $\theta$ stands for the angle between ${\bf r}_{IJ}$ and the horizontal direction in 2D.
%\end{widetext}
% depends on $\theta$, the angle between ${\bf r}_{IJ}$ and the horizontal direction in 2D.
%
%We shall investigate the fate of the topology as a function of structural disorder in the %$M\times\lambda$ phase diagram of the 2DAS model, by varying the band splitting, $M$, and the %interorbital mixing, $\lambda$, while keeping fixed the intraorbital hopping, $t_2$, and the distance %for allowed hoppings $R=|{\bf r}_{IJ}|$. 

The topological phases of the 2DAS model can be classified, as standard, by their topological invariants.
% , for example the Chern number, ${\cal C}$. 
For a pristine square-lattice, the Hamiltonian ${\cal H}$
with nearest neighbors hoppings, $R=
%|{\bf r}_{IJ}|=
a$, can be written in reciprocal space as $H ({\bf k})=d_0({\bf k}){\cal I}+{\bf d}({\bf k})\cdot \boldsymbol{\sigma}$ \cite{SupMat},
where ${\cal I}$ is the identity, $\boldsymbol{\sigma}=(\sigma_x,\sigma_y,\sigma_z)$ are the Pauli matrices,
$d_0({\bf k})=t_2\left[\cos{(k_x a)}+\cos{(k_y a)}\right]$,
and ${\bf d}({\bf k})=(d_x({\bf k}),d_y({\bf k}),d_z({\bf k}))$ with
\begin{eqnarray}
    d_x({\bf k})&=&\lambda\left[1-\cos{(k_x a)}\right]-\sin{(k_x a)}\nonumber\\
    d_y({\bf k})&=&\lambda\left[1-\cos{(k_y a)}\right]-\sin{(k_y a)}\nonumber\\
    d_z({\bf k})&=&(2+M)-\cos{(k_x a)}-\cos{(k_y a)}.
\end{eqnarray}
This defines a map, $T^2\rightarrow X$, between the torus $T^2$ shaped BZ and the target topological space $X$, an image over a Bloch's sphere, $S^2\equiv{\bf d}({\bf k})/|{\bf d}({\bf k})|$, whose degree of map $\pi_2(T^2)\simeq\mathbb{Z}$ is given by the Chern number
\begin{equation}
    {\cal C}(\lambda,M)=\frac{1}{4\pi}
% \iint_{BZ}
    \int_{\rm BZ} \!d^2k\;\hat{\bf d}({\bf k})\cdot
    \left(\nabla_{{\bf k}_x}\hat{\bf d}({\bf k})\times
    \nabla_{{\bf k}_y}\hat{\bf d}({\bf k})\right),
    \label{Chern-Number}
\end{equation}
which is a function of $M$ and $\lambda$, with $\hat{\bf d}({\bf k})={\bf d}({\bf k})/|{\bf d}({\bf k})|$. Since 
% the Chern number 
${\cal C}$ is obtained by an integral over the BZ, its construction relies heavily on translational symmetry. Changes in ${\cal C}(\lambda,M)$ are accompanied by the closing of the gap $\Delta_G({\bf k})\equiv E_+({\bf k})-E_-({\bf k})$, where $E_\pm({\bf k})$
are eigenvalues of $H({\bf k})$.
%$=d_0({\bf k})\pm|{\bf d}({\bf k})|$.

%%%%%%%%%%%%%%%%%%%%%%%%%%%%%%%%%%%%%%%%%%%%%%%%%%%%%%%%%%%%%%%%%%
\begin{figure}
\centering
\includegraphics[width=0.85\columnwidth]{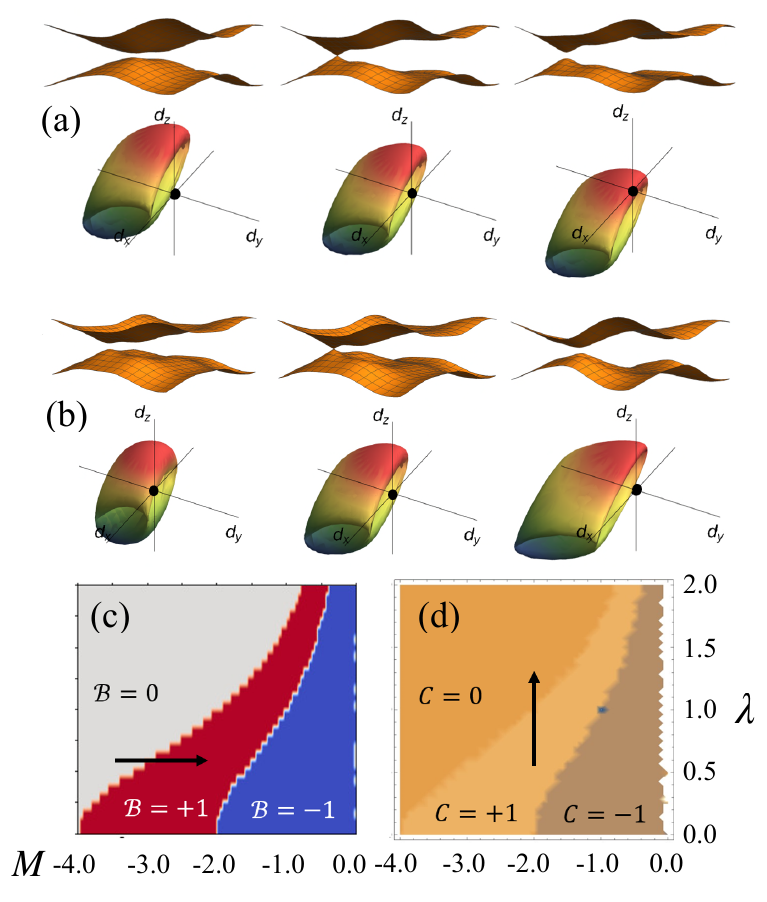}
\caption{Topological transitions in the 2DAS model with $R=1$ and $t_2 = 0.25$. 
(a) Gap $\Delta_G({\bf k})$ and target space ${\bf d}({\bf k})/|{\bf d}({\bf k})|$ for $\lambda=1.0$ and $M=-2.2,-2.0,-1.8$. 
(b) Gap $\Delta_G({\bf k})$ and target space ${\bf d}({\bf k})/|{\bf d}({\bf k})|$ for $M=-2.0$ and $\lambda=0.8,1.0,1.2$. 
The target space colors stand for the value of the Berry curvature at the corresponding ${\bf k}$-point, ranging from red (positive) to green, and blue (negative).
(c) $(M\times\lambda)_{{\cal B}}$ phase diagram and path (arrow) ${\cal B}=0\rightarrow +1$, corresponding to (a).
(d) $(M\times\lambda)_{{\cal C}}$ phase diagram and path (arrow) ${\cal C}=+1\rightarrow 0$, corresponding to (b). The topological markers, ${\cal C}$ and ${\cal B}$, agree perfectly.}
\label{fig:phasediagpristine}
\end{figure}
%%%%%%%%%%%%%%%%%%%%%%%%%%%%%%%%%%%%%%%%%%%%%%%%%%%%%%%%%%%%%%%%%%

Differently from the Chern number, the Bott index ${\cal B}$ is a topological marker defined  directly in coordinate space \cite{Huang2018a, Huang2018b, Loring_2010}. 
To calculate ${\cal B}$ one computes the eigenstates $\left|\psi_i\right>$ of ${\cal H}$
% , Eq.~(\ref{eq:Hamiltonian}), 
with periodic boundary conditions to define the projector of the occupied states $P=\sum_{i}^{N_{\rm occ}}\left|\psi_i\right>\left<\psi_i\right|$. 
Using the sites' positions, one then constructs diagonal matrices representing the position operators $\hat{X}$ and $\hat{Y}$, whose inputs are the remapped coordinates from $[-L,L]$ to $(0,2\pi)$, creating the matrices $\Theta$ and $\Phi$. Finally, one defines the operators, $U=P\exp{(i\Theta)}P$ and $W=P\exp{(i\Phi)}P$, such that 
% the Bott index becomes
%
\begin{equation}
    {\cal B}(\lambda,M)=\frac{1}{2\pi}{\rm Im}\! \left[\mbox{Tr}\big(\log{(WUW^\dagger U^\dagger)}\big)\right].
    \label{Bott-Index}
\end{equation}
${\cal B}$ indicates whether or not the eigenvectors can be reduced to an orthonormal basis of Wannier states. If the system is in a topological state, this reduction is not possible and ${\cal B}\neq 0$ .

Figure~\ref{fig:phasediagpristine} presents the $M\times\lambda$ topological phase diagram 
% obtained via both Eq.~\eqref{Chern-Number} and Eq.~\eqref{Bott-Index}, 
in terms of both ${\cal C}$ and ${\cal B}$ 
for the 2DAS model on a pristine square-lattice with nearest-neighbors hopping. 
In Fig.~\ref{fig:phasediagpristine}(a), for $\lambda=1.0$ and $M=-2.2,-2.0,-1.8$, the target space $X$ is shifted vertically leading to a topological transition when the origin, ${\bf d}({\bf k})=0$, crosses the boundary of $X$ {\it inwards}, entering the (red) region of positive Berry curvature. 
This corresponds to the horizontal arrow in Fig.~\ref{fig:phasediagpristine}(c), such that ${\cal B}=0\rightarrow +1$. 
In Fig.~\ref{fig:phasediagpristine}(b), for $M=-2.0$ and $\lambda=0.8,1.0,1.2$, $X$ is deformed laterally now leading to a topological transition when the origin crosses the boundary of $X$ {\it outwards}, leaving the (red) region of positive Berry curvature. 
This corresponds to ${\cal C}=+1\rightarrow 0$, represented by the vertical arrow in Fig.~\ref{fig:phasediagpristine}(d). 
In both cases, the gap $\Delta_G({\bf k})$ is located along the diagonals in the BZ and closes at the transition as indicated in Figs.~\ref{fig:phasediagpristine}(a) and \ref{fig:phasediagpristine}(b). 
Finally, for ${\cal B}={\cal C}=+1\rightleftharpoons 0\rightleftharpoons -1$ (not shown in the figure), either through variations of $M$ (shift) or $\lambda$ (deformation), the origin inside $X$ crosses over between regions of positive (red) to/from negative (blue) Berry curvatures. 
The gap $\Delta_G({\bf k})$, now located along the ``faces" of the BZ, closes along the critical line, ${\cal B}={\cal C}=0$, and finally reopens and moves towards the $\Gamma$ point as $M\rightarrow 0$, returning the topological markers to a nontrivial value, ${\cal B}={\cal C}=-1$.

\begin{figure}
\centering
\includegraphics[width=0.85\columnwidth]{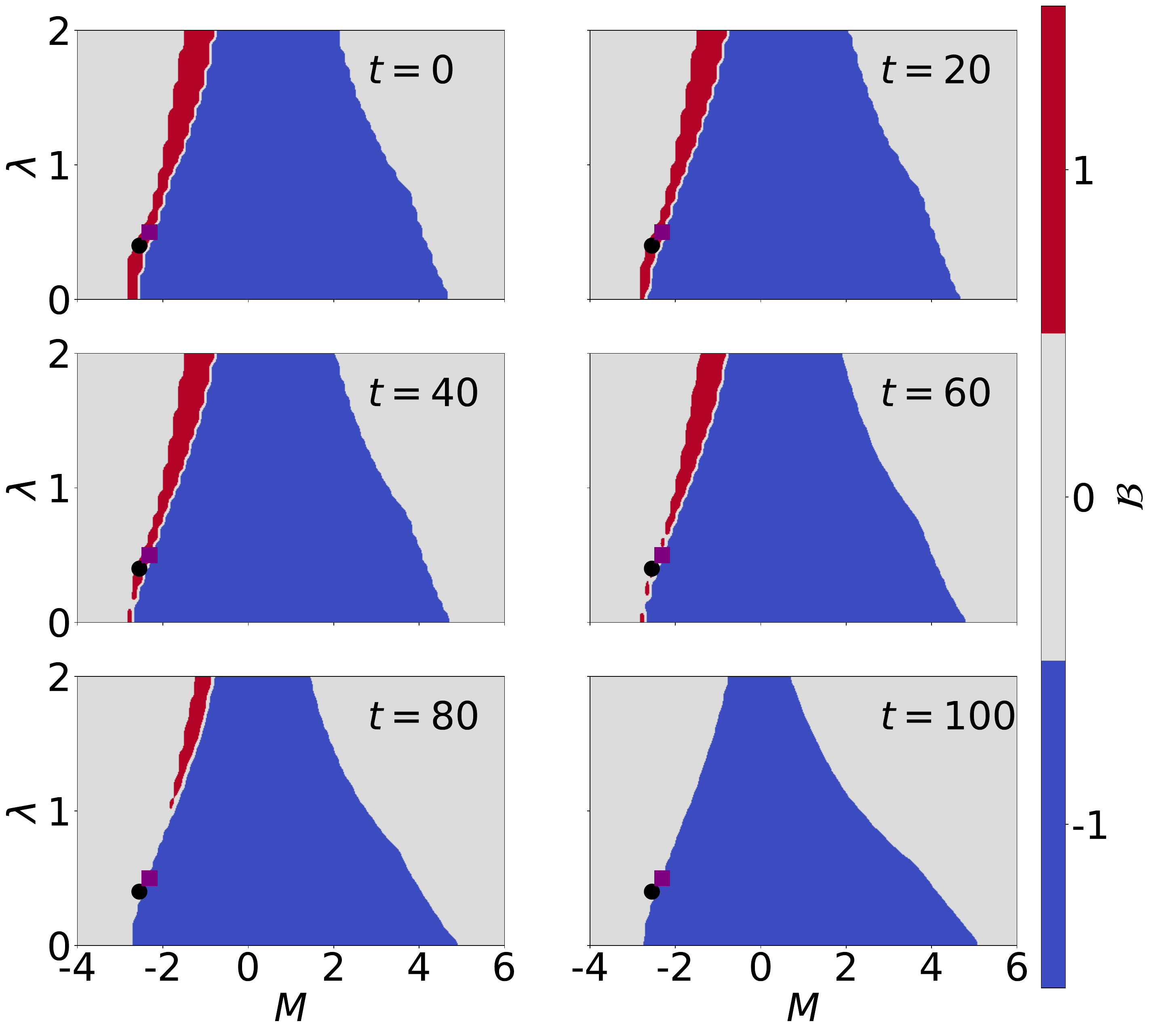}
\caption{Topological phase diagrams $\mathcal{B}(\lambda, M)$ for $R = 4.0$ and $t_2 = 0.25$ and increasing structural disorder:
%represented by the number of iterations in the PN algorithm: 
$t=0$  pristine crystal, $t = 20$ and $t = 40$ paracrystal, $t = 60$ and $t = 80$ amorphous and $t = 100$ random gas. ${\cal B}=+1$ (red), ${\cal B}=-1$ (blue), and ${\cal B}=0$ (gray) regions. 
The topological evolution of the purple square (${\color{violet}\mdblksquare}$) and the black circle (${\color{black}\mdblkcircle}$) points in the $(\lambda, M)$-diagram are discussed in the text.}
\label{fig:phase-diagram}
\end{figure}

%%%%%%%%%%%%%%%%%%%%%%%%%%%%%%%%%%%%%%%%%%%%%%%%%%%%%%%
{\it Algorithm and results.--}
%%%%%%%%%%%%%%%%%%%%%%%%%%%%%%%%%%%%%%%%%%%%%%%%%%%%%%%
Using the PN algorithm for a sequence of up to $t = 100$ iterations, we obtain the structurally-disordered $M\times\lambda$ topological phase diagrams shown in Fig.~\ref{fig:phase-diagram}. 
% The Bott index 
$\mathcal{B}(\lambda, M)$ is computed for ${\cal H}$ 
% of Eq.~\eqref{eq:Hamiltonian} 
on a $24\times 24$ system with periodic boundary conditions at half-filling, with $t_2=0.25$ and $R/a=4$.
The results correspond to $\sim 10^2$ disorder realizations, for which the statistical fluctuations are negligible.  
As the structural disorder increases, with increasing $t$, we observe a gradual but complete suppression of the ${\cal B}=+1$ (red) topological phase. 
We find that this suppression becomes ``slower" with increasing 
% interorbital mixing 
$\lambda$, but the complete destruction of the ${\cal B}=+1$ phase cannot be avoided. 
Contrarily, although the ${\cal B}=-1$ (blue) phase is partially destroyed by structural disorder
%, $t\neq 0$, 
specially for larger values of $\lambda$, a large portion of the ${\cal B}=-1$ phase remains robust all the way into the random gas limit.
% $t=100$. 
For later convenience we stress that, for pristine systems, ${\cal B}=+1$ corresponds to gaps located exclusively away from the $\Gamma$ point, whereas for ${\cal B}=-1$ the gaps are located either at the ``faces" of the BZ or, most importantly, at the $\Gamma$ point. 

%%%%%%%%%%%%%%%%%%%%%%%%%%%%%%%%%%%%%%%%%%%%%%%%%%%%%%%%%%%%%%%%%%%%%%%%%%%%%%%%%%%%%%%%%%
%%%%%%%%%%%%%%%%%%%%%%%%%%%%%%%%%%%%%%%%%%%%%%%%%%%%%%%%%%%%%%%%%%%%%%%%%%%%%%%%%%%%%%%%%%

\begin{figure}
\centering
\includegraphics[width=0.99\columnwidth]{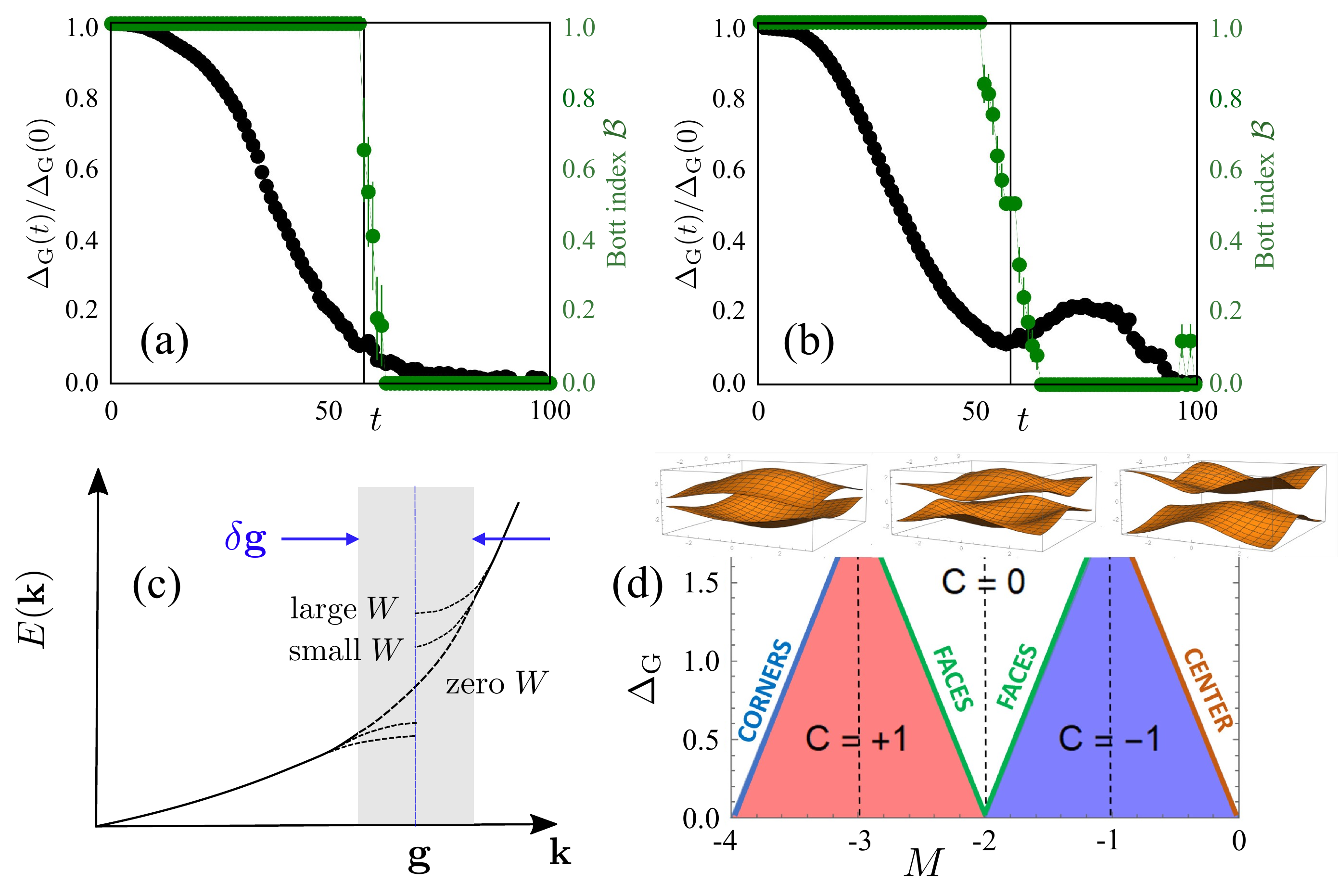}
\caption{Topological gap and Bott index for the (a) black circle $(M = -2.55 \;\; \text{and}\;\; \lambda = 0.4)$ and (b) purple square $(M = -2.31 \;\;\text{and}\;\;\lambda = 0.5)$ of Fig.~\ref{fig:phase-diagram} as a function of structural disorder
%, characterized by the iteration step of the PN algorithm, 
$t$. 
The vertical lines mark the critical $t$ at which ${\cal B}$ changes and the gap closes. 
(c) Dispersion relation at the Bragg planes in first-order pertubation theory.
(d) Sketch of $\Delta_G$ as a function of the band-splitting $M$ in second-order perturbation theory. {\it Inset:} $\Delta_G({\bf k})$ for different $M$ values.}
\label{fig:gap-bott}
\end{figure}

To better understand the evolution of the topological phases 
%in Fig.~\ref{fig:phase-diagram}, 
we show in Fig.~\ref{fig:gap-bott} the average bulk gap, $\Delta_G(t)$, normalized 
by its value at the crystalline phase 
$\Delta_G(t=0)$ and ${\cal B}$ for two representative $(\lambda, M)$ points, the black circle (${\color{black}\mdblkcircle}$) and the purple square (${\color{violet}\mdblksquare}$), indicated in Fig.~\ref{fig:phase-diagram}.
% The point ${\color{black}\mdblkcircle}$ is further away from the ${\cal B}=0$ critical line at $t=0$
Our results show that the gap $\Delta_{G}^{{\color{black}\mdblkcircle}}(t)$ drops quickly with increasing $t$, vanishing at the point of total topological destruction, ${\cal B}=+1 \; (\mbox{red})\rightarrow 0 \; (\mbox{gray})$, remaining closed throughout the remaining steps, see Fig.~\ref{fig:gap-bott}(a). 
Conversely, for the ${\color{violet}\mdblksquare}$-point we find that, although the gap $\Delta_G^{{\color{violet}\mdblksquare}}(t)$ also drops quickly with $t$, it first reaches a minimum at a topological transition, ${\cal B}=+1 \; (\mbox{red})\rightarrow 0 \; (\mbox{gray})\rightarrow -1 \; (\mbox{blue})$, then recovers in the amorphous phase, until it finally closes again in the random limit while preserving topology, ${\cal B}=-1$, see Fig.~\ref{fig:gap-bott}(b). 
Our results suggest that structural disorder provides not only a mechanism for topological destruction, but also a mechanism for disorder induced topological transitions.

%%%%%%%%%%%%%%%%%%%%%%%%%%%%%%%%%%%%%%%%%%%%%%%%%%%%%%%
{\it Topological transitions in paracrystals.--}
%%%%%%%%%%%%%%%%%%%%%%%%%%%%%%%%%%%%%%%%%%%%%%%%%%%%%%%
The interpolation between a pristine crystal and a random gas
%, with disorder, 
was developed by Hosemann \cite{hosemann}. 
Hosemann considered lattice sites position fluctuations of variance $\sigma$
%, that cause pair-correlations to decrease with separation, 
such that the structure factor reads
\begin{equation}
S_\sigma({\bf k}^\prime-{\bf k})=\sum_{{\bf g}}\frac{S_{max}({\bf g})}{1+\ell^2_{hl}({\bf k}^\prime-{\bf k}-{\bf g})^2},
\label{eq:Structure-Factor}
\end{equation}
where ${\bf g}$ stands for the reciprocal lattice vectors.
The amplitudes, $S_{max}({\bf g})=4/\sigma^2{\bf g}^2$, and the breadths for Bragg reflections, $|\delta {\bf g}|\equiv 1/\ell_{hl}=\sigma^2\pi^2(h^2+l^2)/a$, given in terms of the original lattice parameter $a$, allow one to interpolate continuously between pristine and random cases by increasing $\sigma$. 
For $\sigma\rightarrow 0$ we have $\ell_{hl}\rightarrow \infty$ and  $S({\bf k}^\prime-{\bf k})=\sum_{{\bf g}}\delta({\bf k}^\prime-{\bf k}-{\bf g})$, enforcing the kinematic constraints of quasi-momentum conservation at the reciprocal lattice vectors ${\bf g}$ as in a pristine crystal, see Fig.~\ref{Fig:Structures}(a). For $\sigma\rightarrow \infty$ we have $\ell_{hl}\rightarrow 0$ and thus $S({\bf k}^\prime-{\bf k})=S_{max}({\bf 0})\rightarrow 1$, which is isotropic and determined solely by the ${\bf g}=0$ contribution, typical of a completely random lattice,
% and  aperiodic lattice system, 
see Fig.~\ref{Fig:Structures}(d). For $0\leq\sigma\leq\infty$ we have $\infty\geq\ell_{hl}\geq 0$ and $S({\bf k}^\prime-{\bf k})$ is composed by sharp peaks at small ${\bf g}$ (large $\ell_{hl}$) and broader peaks for larger ${\bf g}$ (small $\ell_{hl}$), typical of paracrystals, liquids and amorphous systems, see Figs. \ref{Fig:Structures}(b) and (c). 

%%%%%%%%%%%%%%%%%%%%%%%%%%%%%%%%%%%%%%%%%%%%%%%%%%%%%%%%%%%%%%%%%%%%%%%%%%%%%%%%%%%%%%%%%%
%%%%%%%%%%%%%%%%%%%%%%%%%%%%%%%%%%%%%%%%%%%%%%%%%%%%%%%%%%%%%%%%%%%%%%%%%%%%%%%%%%%%%%%%%%

The evolution of the $\Delta_G(t)$ shown in Fig.~\ref{fig:gap-bott} can now be understood using degenerate perturbation theory for nearly free electrons in a periodic pseudopotential, $W({\bf r})$
% $=W({\bf r}_j)$ 
\cite{harrison1989electronic}. 
To first order, the states of a translationally invariant system at the Bragg planes ${\bf g}$ are ultimately responsible for the opening of a gap of intensity $\Delta_G=|w_{{\bf g}}|$, were $w_{\bf k}$ is the Fourier transform of the pseudopotential.
Structural disorder breaks translational invariance, and the broadening of the structure factor in Eq.~\eqref{eq:Structure-Factor} produces a monotonically decreasing average gap, $\Delta_G=|\overline{w_{{\bf g}}}|$, with disorder average evaluated by including the contributions of all Bragg reflections in an extended zone scheme, each one with its own amplitude $S_{max}({\bf g})$
%= 4/\sigma^2{\bf g}^2$ 
and breadth $|\delta {\bf g}|$, see Fig.~\ref{fig:gap-bott}(c). This is the leading contribution from structural disorder to $\Delta_G^{{\color{black}\mdblkcircle}}(t)$ and also the mechanism promoting the destruction of the ${\cal B}=+1$ phase in Fig.~\ref{fig:phase-diagram}, where all gaps were (initially) located at Bragg planes in the BZ. 

At second order, the corrections to the electronic on-site energies, and consequently to the band-splitting $M$, are caused by variations of the pseudopotential, namely,
\begin{equation}
    \delta M(\sigma)=\alpha_{sp}
    \sum_{{\bf k}-{\bf k}^\prime\neq {\bf g}}S_\sigma({\bf k}-{\bf k}^\prime)
    |w_{{\bf k}-{\bf k}^\prime}|^2 F(|{\bf k}-{\bf k}^\prime|).
\label{eq:DeltaM}
\end{equation}
Here $\alpha_{sp}=(m^*_s-m^*_p)a^2/2\pi\hbar^2$, 
% $w_{{\bf k}-{\bf k}^\prime}$ is the Fourier transform of the pseudopotential, 
$F(|{\bf k}-{\bf k}^\prime|)$ is a positive definite function, and $m^*_s$ and $m^*_p$ the reduced electronic masses for each band \cite{SupMat}. 
For $\alpha_{sp}>0$ the band-splitting $M(\sigma)$ is a monotonic increasing function of the disorder variance $\sigma$, that causes a shift in the the target space $X$ and the occurrence of topological transitions, see Fig.~\ref{fig:gap-bott}(d), whenever the singularity crosses it in between regions of different ${\cal C}$ or ${\cal B}$. 

Let us discuss the effects of structural disorder on the two representative points (${\color{black}\mdblkcircle}$) and (${\color{violet}\mdblksquare}$) initially at the $\mathcal{B} = +1$ topological sector, according to degenerate perturbation theory. To first order, the suppression of Bragg reflections tends to close both $\Delta_{G}^{{\color{black}\mdblkcircle}}(t)$ and $\Delta_G^{{\color{violet}\mdblksquare}}(t)$. 
To second order, the renormalization of $\delta M$ tends to move both $\Delta_{G}^{{\color{black}\mdblkcircle}}(t)$ and $\Delta_G^{{\color{violet}\mdblksquare}}(t)$ towards the $\Gamma$ point. However, the ${\color{violet}\mdblksquare}$-point is closer to the ${\cal B}={\cal C}=0$ critical line than the ${\color{black}\mdblkcircle}$ one. 
Hence, by increasing disorder $\Delta_{G}^{{\color{black}\mdblkcircle}}(t)$ closes before the ${\color{black}\mdblkcircle}$-point escapes the destruction of the $\mathcal{B} = +1$ phase, while the ${\color{violet}\mdblksquare}$-point stays nontrivial and manages to transition to the $\mathcal{B} = -1$ phase, where $\Delta_G^{{\color{violet}\mdblksquare}}(t)$ is sheltered at the $\Gamma$ point. 
This quite remarkable finding shows that the random gas is topological, ${\cal B}=-1$, as a consequence of the fact that the corresponding $\Delta_G^{(\lambda,M)}$ is driven towards the topologically protected $\Gamma$ point, the only ${\bf k}$-point left with finite intensity and breadth, Fig.~\ref{Fig:Structures}(d), after randomization.

%The ${\color{violet}\mdblksquare}$-point  starts ($t=0$) with $\mathcal{B} = +1$ but very close to the ${\cal B}={\cal C}=0$ critical line. As for (${\color{black}\mdblkcircle}$), the first order contribution from structural disorder reduces the gap also for (${\color{violet}\mdblksquare}$), but before it is destroyed, the increase of $\delta M(\sigma)$ in Eq. (\ref{eq:DeltaM}), causes  (${\color{violet}\mdblksquare}$) to undergo a topological transition, $\mathcal{B} = +1 \rightarrow 0$, with the gap closing at the faces of the BZ. Further increase of $\delta M(\sigma)$ causes the gap to reopen, still located at the faces, but now with $\mathcal{B} = -1$, see Fig.~\ref{fig:gap-bott}(d). An even larger increase of $\delta M(\sigma)$ caused by the amorphization, leads the gap to move towards the center of the BZ, the $\Gamma$ point, before the gap is suppressed to zero in the completely random gas limit. Remarkably, however, the random gas limit is topological and has ${\cal B}=-1$, a consequence of the fact that (${\color{violet}\mdblksquare}$) is located at the topologically protected $\Gamma$ point, which is the only point left with nonzero intensity and breadth, see Fig.~\ref{Fig:Structures}(d), after randomization.

%%%%%%%%%%%%%%%%%%%%%%%%%%%%%%%%%%%%%%%%%%%%%%%%%%%%%%%%%
{\it Conclusion.--} 
%%%%%%%%%%%%%%%%%%%%%%%%%%%%%%%%%%%%%%%%%%%%%%%%%%%%%%%%%
Starting from a crystalline topological insulating system, of known topological invariant, $\mathcal{B}\neq 0$, and with gaps located away from the BZ center, we introduce the Perlin noise algorithm to dynamically induce structural disorder in the system, leading it to a completely random state, passing through different intermediate structural phases. 
By considering a two-band model and the generated lattices, we are able to study the occurrence of structure-driven topological phase transitions. We have found numerically that while disorder destroys states corresponding to gaps located at Bragg planes, it also provides a mechanism through which the gaps evolve towards the center of the BZ at the topologically protected $\Gamma$ point. 
Thus, topological transitions can indeed be driven by structural disorder and they can be elegantly described theoretically within Hosemann's paracrystalline theory \cite{hosemann}. 
Our results unveil a scenario that shows the phenomenon of topological protection at its extreme.

%%%%%%%%%%%%%%%%%%%%%%%%%%%%%%%%%%%%%%%%%%%%%%%%%%%%%%%%%
\begin{acknowledgments}
This work is supported by the Brazilian funding agencies FAPERJ and CNPq. 
M.B.S.N. acknowledges the support of FAPESP and the INCT - Materials Informatics.
\end{acknowledgments}
%%%%%%%%%%%%%%%%%%%%%%%%%%%%%%%%%%%%%%%%%%%%%%%%%%%%%%%%%

%%%%%%%%%%%%%%%%%%%%%%%%%%%%%%%%%%%%%%%%%%%%%%%%%%%%%%%%%
\bibliography{amorphousTI}

%merlin.mbs apsrev4-1.bst 2010-07-25 4.21a (PWD, AO, DPC) hacked
%Control: key (0)
%Control: author (8) initials jnrlst
%Control: editor formatted (1) identically to author
%Control: production of article title (-1) disabled
%Control: page (0) single
%Control: year (1) truncated
%Control: production of eprint (0) enabled
\begin{thebibliography}{43}%
\makeatletter
\providecommand \@ifxundefined [1]{%
 \@ifx{#1\undefined}
}%
\providecommand \@ifnum [1]{%
 \ifnum #1\expandafter \@firstoftwo
 \else \expandafter \@secondoftwo
 \fi
}%
\providecommand \@ifx [1]{%
 \ifx #1\expandafter \@firstoftwo
 \else \expandafter \@secondoftwo
 \fi
}%
\providecommand \natexlab [1]{#1}%
\providecommand \enquote  [1]{``#1''}%
\providecommand \bibnamefont  [1]{#1}%
\providecommand \bibfnamefont [1]{#1}%
\providecommand \citenamefont [1]{#1}%
\providecommand \href@noop [0]{\@secondoftwo}%
\providecommand \href [0]{\begingroup \@sanitize@url \@href}%
\providecommand \@href[1]{\@@startlink{#1}\@@href}%
\providecommand \@@href[1]{\endgroup#1\@@endlink}%
\providecommand \@sanitize@url [0]{\catcode `\\12\catcode `\$12\catcode
  `\&12\catcode `\#12\catcode `\^12\catcode `\_12\catcode `\%12\relax}%
\providecommand \@@startlink[1]{}%
\providecommand \@@endlink[0]{}%
\providecommand \url  [0]{\begingroup\@sanitize@url \@url }%
\providecommand \@url [1]{\endgroup\@href {#1}{\urlprefix }}%
\providecommand \urlprefix  [0]{URL }%
\providecommand \Eprint [0]{\href }%
\providecommand \doibase [0]{http://dx.doi.org/}%
\providecommand \selectlanguage [0]{\@gobble}%
\providecommand \bibinfo  [0]{\@secondoftwo}%
\providecommand \bibfield  [0]{\@secondoftwo}%
\providecommand \translation [1]{[#1]}%
\providecommand \BibitemOpen [0]{}%
\providecommand \bibitemStop [0]{}%
\providecommand \bibitemNoStop [0]{.\EOS\space}%
\providecommand \EOS [0]{\spacefactor3000\relax}%
\providecommand \BibitemShut  [1]{\csname bibitem#1\endcsname}%
\let\auto@bib@innerbib\@empty
%</preamble>
\bibitem [{\citenamefont {Hasan}\ and\ \citenamefont {Kane}(2010)}]{Hasan2010}%
  \BibitemOpen
  \bibfield  {author} {\bibinfo {author} {\bibfnamefont {M.~Z.}\ \bibnamefont
  {Hasan}}\ and\ \bibinfo {author} {\bibfnamefont {C.~L.}\ \bibnamefont
  {Kane}},\ }\href {\doibase 10.1103/RevModPhys.82.3045} {\bibfield  {journal}
  {\bibinfo  {journal} {Rev. Mod. Phys.}\ }\textbf {\bibinfo {volume} {82}},\
  \bibinfo {pages} {3045} (\bibinfo {year} {2010})}\BibitemShut {NoStop}%
\bibitem [{\citenamefont {Qi}\ and\ \citenamefont {Zhang}(2011)}]{Qi2011}%
  \BibitemOpen
  \bibfield  {author} {\bibinfo {author} {\bibfnamefont {X.-L.}\ \bibnamefont
  {Qi}}\ and\ \bibinfo {author} {\bibfnamefont {S.-C.}\ \bibnamefont {Zhang}},\
  }\href {\doibase 10.1103/RevModPhys.83.1057} {\bibfield  {journal} {\bibinfo
  {journal} {Rev. Mod. Phys.}\ }\textbf {\bibinfo {volume} {83}},\ \bibinfo
  {pages} {1057} (\bibinfo {year} {2011})}\BibitemShut {NoStop}%
\bibitem [{\citenamefont {Thouless}\ \emph {et~al.}(1982)\citenamefont
  {Thouless}, \citenamefont {Kohmoto}, \citenamefont {Nightingale},\ and\
  \citenamefont {{den Nijs}}}]{TKNN}%
  \BibitemOpen
  \bibfield  {author} {\bibinfo {author} {\bibfnamefont {D.~J.}\ \bibnamefont
  {Thouless}}, \bibinfo {author} {\bibfnamefont {M.}~\bibnamefont {Kohmoto}},
  \bibinfo {author} {\bibfnamefont {M.~P.}\ \bibnamefont {Nightingale}}, \ and\
  \bibinfo {author} {\bibfnamefont {M.}~\bibnamefont {{den Nijs}}},\ }\href
  {\doibase 10.1103/PhysRevLett.49.405} {\bibfield  {journal} {\bibinfo
  {journal} {Phys. Rev. Lett.}\ }\textbf {\bibinfo {volume} {49}},\ \bibinfo
  {pages} {405} (\bibinfo {year} {1982})}\BibitemShut {NoStop}%
\bibitem [{\citenamefont {Hatsugai}(1993)}]{Hatsugai1990-1}%
  \BibitemOpen
  \bibfield  {author} {\bibinfo {author} {\bibfnamefont {Y.}~\bibnamefont
  {Hatsugai}},\ }\href {\doibase 10.1103/PhysRevLett.71.3697} {\bibfield
  {journal} {\bibinfo  {journal} {Phys. Rev. Lett.}\ }\textbf {\bibinfo
  {volume} {71}},\ \bibinfo {pages} {3697} (\bibinfo {year}
  {1993})}\BibitemShut {NoStop}%
\bibitem [{\citenamefont {Bernevig}(2013)}]{bernevig}%
  \BibitemOpen
  \bibfield  {author} {\bibinfo {author} {\bibfnamefont {B.~A.}\ \bibnamefont
  {Bernevig}},\ }\href {\doibase doi:10.1515/9781400846733} {\emph {\bibinfo
  {title} {Topological Insulators and Topological Superconductors}}}\ (\bibinfo
   {publisher} {Princeton University Press},\ \bibinfo {address} {Princeton},\
  \bibinfo {year} {2013})\BibitemShut {NoStop}%
\bibitem [{\citenamefont {Kane}\ and\ \citenamefont
  {Mele}(2005{\natexlab{a}})}]{kane-mele}%
  \BibitemOpen
  \bibfield  {author} {\bibinfo {author} {\bibfnamefont {C.~L.}\ \bibnamefont
  {Kane}}\ and\ \bibinfo {author} {\bibfnamefont {E.~J.}\ \bibnamefont
  {Mele}},\ }\href {\doibase 10.1103/PhysRevLett.95.226801} {\bibfield
  {journal} {\bibinfo  {journal} {Phys. Rev. Lett.}\ }\textbf {\bibinfo
  {volume} {95}},\ \bibinfo {pages} {226801} (\bibinfo {year}
  {2005}{\natexlab{a}})}\BibitemShut {NoStop}%
\bibitem [{\citenamefont {Kane}\ and\ \citenamefont
  {Mele}(2005{\natexlab{b}})}]{kane-mele-z2}%
  \BibitemOpen
  \bibfield  {author} {\bibinfo {author} {\bibfnamefont {C.~L.}\ \bibnamefont
  {Kane}}\ and\ \bibinfo {author} {\bibfnamefont {E.~J.}\ \bibnamefont
  {Mele}},\ }\href {\doibase 10.1103/PhysRevLett.95.146802} {\bibfield
  {journal} {\bibinfo  {journal} {Phys. Rev. Lett.}\ }\textbf {\bibinfo
  {volume} {95}},\ \bibinfo {pages} {146802} (\bibinfo {year}
  {2005}{\natexlab{b}})}\BibitemShut {NoStop}%
\bibitem [{\citenamefont {Bernevig}\ \emph {et~al.}(2006)\citenamefont
  {Bernevig}, \citenamefont {Hughes},\ and\ \citenamefont {Zhang}}]{bhz}%
  \BibitemOpen
  \bibfield  {author} {\bibinfo {author} {\bibfnamefont {B.~A.}\ \bibnamefont
  {Bernevig}}, \bibinfo {author} {\bibfnamefont {T.~L.}\ \bibnamefont
  {Hughes}}, \ and\ \bibinfo {author} {\bibfnamefont {S.-C.}\ \bibnamefont
  {Zhang}},\ }\href {\doibase 10.1126/science.1133734} {\bibfield  {journal}
  {\bibinfo  {journal} {Science}\ }\textbf {\bibinfo {volume} {314}},\ \bibinfo
  {pages} {1757} (\bibinfo {year} {2006})}\BibitemShut {NoStop}%
\bibitem [{\citenamefont {K{\"o}nig}\ \emph {et~al.}(2007)\citenamefont
  {K{\"o}nig}, \citenamefont {Wiedmann}, \citenamefont {Br{\"u}ne},
  \citenamefont {Roth}, \citenamefont {Buhmann}, \citenamefont {Molenkamp},
  \citenamefont {Qi},\ and\ \citenamefont {Zhang}}]{Konig2007}%
  \BibitemOpen
  \bibfield  {author} {\bibinfo {author} {\bibfnamefont {M.}~\bibnamefont
  {K{\"o}nig}}, \bibinfo {author} {\bibfnamefont {S.}~\bibnamefont {Wiedmann}},
  \bibinfo {author} {\bibfnamefont {C.}~\bibnamefont {Br{\"u}ne}}, \bibinfo
  {author} {\bibfnamefont {A.}~\bibnamefont {Roth}}, \bibinfo {author}
  {\bibfnamefont {H.}~\bibnamefont {Buhmann}}, \bibinfo {author} {\bibfnamefont
  {L.~W.}\ \bibnamefont {Molenkamp}}, \bibinfo {author} {\bibfnamefont {X.-L.}\
  \bibnamefont {Qi}}, \ and\ \bibinfo {author} {\bibfnamefont {S.-C.}\
  \bibnamefont {Zhang}},\ }\href {\doibase 10.1126/science.1148047} {\bibfield
  {journal} {\bibinfo  {journal} {Science}\ }\textbf {\bibinfo {volume}
  {318}},\ \bibinfo {pages} {766} (\bibinfo {year} {2007})}\BibitemShut
  {NoStop}%
\bibitem [{\citenamefont {Roy}(2009)}]{Roy2009}%
  \BibitemOpen
  \bibfield  {author} {\bibinfo {author} {\bibfnamefont {R.}~\bibnamefont
  {Roy}},\ }\href {\doibase 10.1103/PhysRevB.79.195322} {\bibfield  {journal}
  {\bibinfo  {journal} {Phys. Rev. B}\ }\textbf {\bibinfo {volume} {79}},\
  \bibinfo {pages} {195322} (\bibinfo {year} {2009})}\BibitemShut {NoStop}%
\bibitem [{\citenamefont {Kohmoto}(1985)}]{Kohmoto1985}%
  \BibitemOpen
  \bibfield  {author} {\bibinfo {author} {\bibfnamefont {M.}~\bibnamefont
  {Kohmoto}},\ }\href {\doibase https://doi.org/10.1016/0003-4916(85)90148-4}
  {\bibfield  {journal} {\bibinfo  {journal} {Ann. Phys. (N. Y.)}\ }\textbf
  {\bibinfo {volume} {160}},\ \bibinfo {pages} {343} (\bibinfo {year}
  {1985})}\BibitemShut {NoStop}%
\bibitem [{\citenamefont {Qi}\ \emph {et~al.}(2008)\citenamefont {Qi},
  \citenamefont {Hughes},\ and\ \citenamefont {Zhang}}]{Qi2008}%
  \BibitemOpen
  \bibfield  {author} {\bibinfo {author} {\bibfnamefont {X.-L.}\ \bibnamefont
  {Qi}}, \bibinfo {author} {\bibfnamefont {T.~L.}\ \bibnamefont {Hughes}}, \
  and\ \bibinfo {author} {\bibfnamefont {S.-C.}\ \bibnamefont {Zhang}},\ }\href
  {\doibase 10.1103/PhysRevB.78.195424} {\bibfield  {journal} {\bibinfo
  {journal} {Phys. Rev. B}\ }\textbf {\bibinfo {volume} {78}},\ \bibinfo
  {pages} {195424} (\bibinfo {year} {2008})}\BibitemShut {NoStop}%
\bibitem [{\citenamefont {Moessner}\ and\ \citenamefont
  {Moore}(2021)}]{Moessner2021}%
  \BibitemOpen
  \bibfield  {author} {\bibinfo {author} {\bibfnamefont {R.}~\bibnamefont
  {Moessner}}\ and\ \bibinfo {author} {\bibfnamefont {J.}~\bibnamefont
  {Moore}},\ }\href {https://books.google.com.br/books?id=QBv2zQEACAAJ} {\emph
  {\bibinfo {title} {Topological Phases of Matter}}}\ (\bibinfo  {publisher}
  {Cambridge University Press},\ \bibinfo {year} {2021})\BibitemShut {NoStop}%
\bibitem [{\citenamefont {Li}\ \emph {et~al.}(2009)\citenamefont {Li},
  \citenamefont {Chu}, \citenamefont {Jain},\ and\ \citenamefont
  {Shen}}]{Li2009}%
  \BibitemOpen
  \bibfield  {author} {\bibinfo {author} {\bibfnamefont {J.}~\bibnamefont
  {Li}}, \bibinfo {author} {\bibfnamefont {R.-L.}\ \bibnamefont {Chu}},
  \bibinfo {author} {\bibfnamefont {J.~K.}\ \bibnamefont {Jain}}, \ and\
  \bibinfo {author} {\bibfnamefont {S.-Q.}\ \bibnamefont {Shen}},\ }\href
  {\doibase 10.1103/PhysRevLett.102.136806} {\bibfield  {journal} {\bibinfo
  {journal} {Phys. Rev. Lett.}\ }\textbf {\bibinfo {volume} {102}},\ \bibinfo
  {pages} {136806} (\bibinfo {year} {2009})}\BibitemShut {NoStop}%
\bibitem [{\citenamefont {Groth}\ \emph {et~al.}(2009)\citenamefont {Groth},
  \citenamefont {Wimmer}, \citenamefont {Akhmerov}, \citenamefont
  {Tworzyd\l{}o},\ and\ \citenamefont {Beenakker}}]{Groth2009}%
  \BibitemOpen
  \bibfield  {author} {\bibinfo {author} {\bibfnamefont {C.~W.}\ \bibnamefont
  {Groth}}, \bibinfo {author} {\bibfnamefont {M.}~\bibnamefont {Wimmer}},
  \bibinfo {author} {\bibfnamefont {A.~R.}\ \bibnamefont {Akhmerov}}, \bibinfo
  {author} {\bibfnamefont {J.}~\bibnamefont {Tworzyd\l{}o}}, \ and\ \bibinfo
  {author} {\bibfnamefont {C.~W.~J.}\ \bibnamefont {Beenakker}},\ }\href
  {\doibase 10.1103/PhysRevLett.103.196805} {\bibfield  {journal} {\bibinfo
  {journal} {Phys. Rev. Lett.}\ }\textbf {\bibinfo {volume} {103}},\ \bibinfo
  {pages} {196805} (\bibinfo {year} {2009})}\BibitemShut {NoStop}%
\bibitem [{\citenamefont {Agarwala}\ and\ \citenamefont
  {Shenoy}(2017)}]{Agarwala2017}%
  \BibitemOpen
  \bibfield  {author} {\bibinfo {author} {\bibfnamefont {A.}~\bibnamefont
  {Agarwala}}\ and\ \bibinfo {author} {\bibfnamefont {V.~B.}\ \bibnamefont
  {Shenoy}},\ }\href {\doibase 10.1103/PhysRevLett.118.236402} {\bibfield
  {journal} {\bibinfo  {journal} {Phys. Rev. Lett.}\ }\textbf {\bibinfo
  {volume} {118}},\ \bibinfo {pages} {236402} (\bibinfo {year}
  {2017})}\BibitemShut {NoStop}%
\bibitem [{\citenamefont {Mitchell}\ \emph {et~al.}(2018)\citenamefont
  {Mitchell}, \citenamefont {Nash}, \citenamefont {Hexner}, \citenamefont
  {Turner},\ and\ \citenamefont {Irvine}}]{Mitchell2018}%
  \BibitemOpen
  \bibfield  {author} {\bibinfo {author} {\bibfnamefont {N.~P.}\ \bibnamefont
  {Mitchell}}, \bibinfo {author} {\bibfnamefont {L.~M.}\ \bibnamefont {Nash}},
  \bibinfo {author} {\bibfnamefont {D.}~\bibnamefont {Hexner}}, \bibinfo
  {author} {\bibfnamefont {A.~M.}\ \bibnamefont {Turner}}, \ and\ \bibinfo
  {author} {\bibfnamefont {W.~T.~M.}\ \bibnamefont {Irvine}},\ }\href {\doibase
  10.1038/s41567-017-0024-5} {\bibfield  {journal} {\bibinfo  {journal} {Nat.
  Phys.}\ }\textbf {\bibinfo {volume} {18}},\ \bibinfo {pages} {380} (\bibinfo
  {year} {2018})}\BibitemShut {NoStop}%
\bibitem [{\citenamefont {Varjas}\ \emph {et~al.}(2019)\citenamefont {Varjas},
  \citenamefont {Lau}, \citenamefont {P\"oyh\"onen}, \citenamefont {Akhmerov},
  \citenamefont {Pikulin},\ and\ \citenamefont {Fulga}}]{Varjas2019}%
  \BibitemOpen
  \bibfield  {author} {\bibinfo {author} {\bibfnamefont {D.}~\bibnamefont
  {Varjas}}, \bibinfo {author} {\bibfnamefont {A.}~\bibnamefont {Lau}},
  \bibinfo {author} {\bibfnamefont {K.}~\bibnamefont {P\"oyh\"onen}}, \bibinfo
  {author} {\bibfnamefont {A.~R.}\ \bibnamefont {Akhmerov}}, \bibinfo {author}
  {\bibfnamefont {D.~I.}\ \bibnamefont {Pikulin}}, \ and\ \bibinfo {author}
  {\bibfnamefont {I.~C.}\ \bibnamefont {Fulga}},\ }\href {\doibase
  10.1103/PhysRevLett.123.196401} {\bibfield  {journal} {\bibinfo  {journal}
  {Phys. Rev. Lett.}\ }\textbf {\bibinfo {volume} {123}},\ \bibinfo {pages}
  {196401} (\bibinfo {year} {2019})}\BibitemShut {NoStop}%
\bibitem [{\citenamefont {Yang}\ \emph {et~al.}(2019)\citenamefont {Yang},
  \citenamefont {Qin}, \citenamefont {Deng}, \citenamefont {Duan},\ and\
  \citenamefont {Xu}}]{Yang2019}%
  \BibitemOpen
  \bibfield  {author} {\bibinfo {author} {\bibfnamefont {Y.-B.}\ \bibnamefont
  {Yang}}, \bibinfo {author} {\bibfnamefont {T.}~\bibnamefont {Qin}}, \bibinfo
  {author} {\bibfnamefont {D.-L.}\ \bibnamefont {Deng}}, \bibinfo {author}
  {\bibfnamefont {L.-M.}\ \bibnamefont {Duan}}, \ and\ \bibinfo {author}
  {\bibfnamefont {Y.}~\bibnamefont {Xu}},\ }\href {\doibase
  10.1103/PhysRevLett.123.076401} {\bibfield  {journal} {\bibinfo  {journal}
  {Phys. Rev. Lett.}\ }\textbf {\bibinfo {volume} {123}},\ \bibinfo {pages}
  {076401} (\bibinfo {year} {2019})}\BibitemShut {NoStop}%
\bibitem [{\citenamefont {Huang}\ and\ \citenamefont
  {Liu}(2018{\natexlab{a}})}]{Huang2018a}%
  \BibitemOpen
  \bibfield  {author} {\bibinfo {author} {\bibfnamefont {H.}~\bibnamefont
  {Huang}}\ and\ \bibinfo {author} {\bibfnamefont {F.}~\bibnamefont {Liu}},\
  }\href {\doibase 10.1103/PhysRevLett.121.126401} {\bibfield  {journal}
  {\bibinfo  {journal} {Phys. Rev. Lett.}\ }\textbf {\bibinfo {volume} {121}},\
  \bibinfo {pages} {126401} (\bibinfo {year} {2018}{\natexlab{a}})}\BibitemShut
  {NoStop}%
\bibitem [{\citenamefont {Huang}\ and\ \citenamefont
  {Liu}(2018{\natexlab{b}})}]{Huang2018b}%
  \BibitemOpen
  \bibfield  {author} {\bibinfo {author} {\bibfnamefont {H.}~\bibnamefont
  {Huang}}\ and\ \bibinfo {author} {\bibfnamefont {F.}~\bibnamefont {Liu}},\
  }\href {\doibase 10.1103/PhysRevB.98.125130} {\bibfield  {journal} {\bibinfo
  {journal} {Phys. Rev. B}\ }\textbf {\bibinfo {volume} {98}},\ \bibinfo
  {pages} {125130} (\bibinfo {year} {2018}{\natexlab{b}})}\BibitemShut
  {NoStop}%
\bibitem [{\citenamefont {Marsal}\ \emph {et~al.}(2020)\citenamefont {Marsal},
  \citenamefont {Varjas},\ and\ \citenamefont {Grushin}}]{Marsal2020}%
  \BibitemOpen
  \bibfield  {author} {\bibinfo {author} {\bibfnamefont {Q.}~\bibnamefont
  {Marsal}}, \bibinfo {author} {\bibfnamefont {D.}~\bibnamefont {Varjas}}, \
  and\ \bibinfo {author} {\bibfnamefont {A.~G.}\ \bibnamefont {Grushin}},\
  }\href {\doibase 10.1073/pnas.2007384117} {\bibfield  {journal} {\bibinfo
  {journal} {Proc. Natl. Acad. Sci. U.S.A.}\ }\textbf {\bibinfo {volume}
  {117}},\ \bibinfo {pages} {30260} (\bibinfo {year} {2020})}\BibitemShut
  {NoStop}%
\bibitem [{\citenamefont {Agarwala}\ \emph {et~al.}(2020)\citenamefont
  {Agarwala}, \citenamefont {Juri\ifmmode \check{c}\else
  \v{c}\fi{}i\ifmmode~\acute{c}\else \'{c}\fi{}},\ and\ \citenamefont
  {Roy}}]{Agarwala2020}%
  \BibitemOpen
  \bibfield  {author} {\bibinfo {author} {\bibfnamefont {A.}~\bibnamefont
  {Agarwala}}, \bibinfo {author} {\bibfnamefont {V.}~\bibnamefont {Juri\ifmmode
  \check{c}\else \v{c}\fi{}i\ifmmode~\acute{c}\else \'{c}\fi{}}}, \ and\
  \bibinfo {author} {\bibfnamefont {B.}~\bibnamefont {Roy}},\ }\href {\doibase
  10.1103/PhysRevResearch.2.012067} {\bibfield  {journal} {\bibinfo  {journal}
  {Phys. Rev. Res.}\ }\textbf {\bibinfo {volume} {2}},\ \bibinfo {pages}
  {012067} (\bibinfo {year} {2020})}\BibitemShut {NoStop}%
\bibitem [{\citenamefont {Corbae}\ \emph
  {et~al.}(2023{\natexlab{a}})\citenamefont {Corbae}, \citenamefont
  {Hannukainen}, \citenamefont {Marsal}, \citenamefont {Muñoz-Segovia},\ and\
  \citenamefont {Grushin}}]{Corbae2023b}%
  \BibitemOpen
  \bibfield  {author} {\bibinfo {author} {\bibfnamefont {P.}~\bibnamefont
  {Corbae}}, \bibinfo {author} {\bibfnamefont {J.~D.}\ \bibnamefont
  {Hannukainen}}, \bibinfo {author} {\bibfnamefont {Q.}~\bibnamefont {Marsal}},
  \bibinfo {author} {\bibfnamefont {D.}~\bibnamefont {Muñoz-Segovia}}, \ and\
  \bibinfo {author} {\bibfnamefont {A.~G.}\ \bibnamefont {Grushin}},\ }\href
  {\doibase 10.1209/0295-5075/acc2e2} {\bibfield  {journal} {\bibinfo
  {journal} {EPL}\ }\textbf {\bibinfo {volume} {142}},\ \bibinfo {pages}
  {16001} (\bibinfo {year} {2023}{\natexlab{a}})}\BibitemShut {NoStop}%
\bibitem [{\citenamefont {Costa}\ \emph {et~al.}(2019)\citenamefont {Costa},
  \citenamefont {Schleder}, \citenamefont {Nardelli}, \citenamefont
  {Lewenkopf},\ and\ \citenamefont {Fazzio}}]{Costa2019}%
  \BibitemOpen
  \bibfield  {author} {\bibinfo {author} {\bibfnamefont {M.}~\bibnamefont
  {Costa}}, \bibinfo {author} {\bibfnamefont {G.~R.}\ \bibnamefont {Schleder}},
  \bibinfo {author} {\bibfnamefont {M.~B.}\ \bibnamefont {Nardelli}}, \bibinfo
  {author} {\bibfnamefont {C.}~\bibnamefont {Lewenkopf}}, \ and\ \bibinfo
  {author} {\bibfnamefont {A.}~\bibnamefont {Fazzio}},\ }\href {\doibase
  10.1021/acs.nanolett.9b03881} {\bibfield  {journal} {\bibinfo  {journal}
  {Nano Lett.}\ }\textbf {\bibinfo {volume} {19}},\ \bibinfo {pages} {8941}
  (\bibinfo {year} {2019})}\BibitemShut {NoStop}%
\bibitem [{\citenamefont {Focassio}\ \emph
  {et~al.}(2021{\natexlab{a}})\citenamefont {Focassio}, \citenamefont
  {Schleder}, \citenamefont {Costa}, \citenamefont {Fazzio},\ and\
  \citenamefont {Lewenkopf}}]{Focassio2021a}%
  \BibitemOpen
  \bibfield  {author} {\bibinfo {author} {\bibfnamefont {B.}~\bibnamefont
  {Focassio}}, \bibinfo {author} {\bibfnamefont {G.~R.}\ \bibnamefont
  {Schleder}}, \bibinfo {author} {\bibfnamefont {M.}~\bibnamefont {Costa}},
  \bibinfo {author} {\bibfnamefont {A.}~\bibnamefont {Fazzio}}, \ and\ \bibinfo
  {author} {\bibfnamefont {C.}~\bibnamefont {Lewenkopf}},\ }\href {\doibase
  10.1088/2053-1583/abdb97} {\bibfield  {journal} {\bibinfo  {journal} {2D
  Mater.}\ }\textbf {\bibinfo {volume} {8}},\ \bibinfo {pages} {025032}
  (\bibinfo {year} {2021}{\natexlab{a}})}\BibitemShut {NoStop}%
\bibitem [{\citenamefont {Corbae}\ \emph
  {et~al.}(2023{\natexlab{b}})\citenamefont {Corbae}, \citenamefont {Ciocys},
  \citenamefont {Varjas}, \citenamefont {Kennedy}, \citenamefont {Zeltmann},
  \citenamefont {Molina-Ruiz}, \citenamefont {Griffin}, \citenamefont
  {Jozwiak}, \citenamefont {Chen}, \citenamefont {Wang}, \citenamefont {Minor},
  \citenamefont {Scott}, \citenamefont {Grushin}, \citenamefont {Lanzara},\
  and\ \citenamefont {Hellman}}]{Corbae2023a}%
  \BibitemOpen
  \bibfield  {author} {\bibinfo {author} {\bibfnamefont {P.}~\bibnamefont
  {Corbae}}, \bibinfo {author} {\bibfnamefont {S.}~\bibnamefont {Ciocys}},
  \bibinfo {author} {\bibfnamefont {D.}~\bibnamefont {Varjas}}, \bibinfo
  {author} {\bibfnamefont {E.}~\bibnamefont {Kennedy}}, \bibinfo {author}
  {\bibfnamefont {S.}~\bibnamefont {Zeltmann}}, \bibinfo {author}
  {\bibfnamefont {M.}~\bibnamefont {Molina-Ruiz}}, \bibinfo {author}
  {\bibfnamefont {S.~M.}\ \bibnamefont {Griffin}}, \bibinfo {author}
  {\bibfnamefont {C.}~\bibnamefont {Jozwiak}}, \bibinfo {author} {\bibfnamefont
  {Z.}~\bibnamefont {Chen}}, \bibinfo {author} {\bibfnamefont {L.-W.}\
  \bibnamefont {Wang}}, \bibinfo {author} {\bibfnamefont {A.~M.}\ \bibnamefont
  {Minor}}, \bibinfo {author} {\bibfnamefont {M.}~\bibnamefont {Scott}},
  \bibinfo {author} {\bibfnamefont {A.~G.}\ \bibnamefont {Grushin}}, \bibinfo
  {author} {\bibfnamefont {A.}~\bibnamefont {Lanzara}}, \ and\ \bibinfo
  {author} {\bibfnamefont {F.}~\bibnamefont {Hellman}},\ }\href {\doibase
  10.1038/s41563-022-01458-0} {\bibfield  {journal} {\bibinfo  {journal} {Nat.
  Mater.}\ }\textbf {\bibinfo {volume} {22}},\ \bibinfo {pages} {200} (\bibinfo
  {year} {2023}{\natexlab{b}})}\BibitemShut {NoStop}%
\bibitem [{\citenamefont {Hosemann}\ and\ \citenamefont
  {Hindeleh}(1995)}]{hosemann}%
  \BibitemOpen
  \bibfield  {author} {\bibinfo {author} {\bibfnamefont {R.}~\bibnamefont
  {Hosemann}}\ and\ \bibinfo {author} {\bibfnamefont {A.~M.}\ \bibnamefont
  {Hindeleh}},\ }\href {\doibase 10.1080/00222349508219497} {\bibfield
  {journal} {\bibinfo  {journal} {Journal of Macromolecular Science, Part B}\
  }\textbf {\bibinfo {volume} {34}},\ \bibinfo {pages} {327} (\bibinfo {year}
  {1995})}\BibitemShut {NoStop}%
\bibitem [{\citenamefont {Sahlberg}\ \emph {et~al.}(2020)\citenamefont
  {Sahlberg}, \citenamefont {Weststr\"om}, \citenamefont {P\"oyh\"onen},\ and\
  \citenamefont {Ojanen}}]{Sahlberg2020}%
  \BibitemOpen
  \bibfield  {author} {\bibinfo {author} {\bibfnamefont {I.}~\bibnamefont
  {Sahlberg}}, \bibinfo {author} {\bibfnamefont {A.}~\bibnamefont
  {Weststr\"om}}, \bibinfo {author} {\bibfnamefont {K.}~\bibnamefont
  {P\"oyh\"onen}}, \ and\ \bibinfo {author} {\bibfnamefont {T.}~\bibnamefont
  {Ojanen}},\ }\href {\doibase 10.1103/PhysRevResearch.2.013053} {\bibfield
  {journal} {\bibinfo  {journal} {Phys. Rev. Res.}\ }\textbf {\bibinfo {volume}
  {2}},\ \bibinfo {pages} {013053} (\bibinfo {year} {2020})}\BibitemShut
  {NoStop}%
\bibitem [{\citenamefont {Florescu}\ \emph {et~al.}(2009)\citenamefont
  {Florescu}, \citenamefont {Torquato},\ and\ \citenamefont
  {Steinhardt}}]{Florescu2009}%
  \BibitemOpen
  \bibfield  {author} {\bibinfo {author} {\bibfnamefont {M.}~\bibnamefont
  {Florescu}}, \bibinfo {author} {\bibfnamefont {S.}~\bibnamefont {Torquato}},
  \ and\ \bibinfo {author} {\bibfnamefont {P.~J.}\ \bibnamefont {Steinhardt}},\
  }\href {\doibase 10.1073/pnas.0907744106} {\bibfield  {journal} {\bibinfo
  {journal} {Proc. Natl. Acad. Sci. U.S.A.}\ }\textbf {\bibinfo {volume}
  {106}},\ \bibinfo {pages} {20658} (\bibinfo {year} {2009})}\BibitemShut
  {NoStop}%
\bibitem [{\citenamefont {Klatt}\ \emph {et~al.}(2019)\citenamefont {Klatt},
  \citenamefont {Lovri\'c}, \citenamefont {Chen}, \citenamefont {Kapfer},
  \citenamefont {Schaller}, \citenamefont {Sch\"onh\"ofer}, \citenamefont
  {Gardiner}, \citenamefont {Ana-Sun\v{c}ana~Smith},\ and\ \citenamefont
  {Torquato}}]{Klatt2018}%
  \BibitemOpen
  \bibfield  {author} {\bibinfo {author} {\bibfnamefont {M.~A.}\ \bibnamefont
  {Klatt}}, \bibinfo {author} {\bibfnamefont {J.}~\bibnamefont {Lovri\'c}},
  \bibinfo {author} {\bibfnamefont {D.}~\bibnamefont {Chen}}, \bibinfo {author}
  {\bibfnamefont {S.~C.}\ \bibnamefont {Kapfer}}, \bibinfo {author}
  {\bibfnamefont {F.~M.}\ \bibnamefont {Schaller}}, \bibinfo {author}
  {\bibfnamefont {P.~W.~A.}\ \bibnamefont {Sch\"onh\"ofer}}, \bibinfo {author}
  {\bibfnamefont {B.~S.}\ \bibnamefont {Gardiner}}, \bibinfo {author}
  {\bibfnamefont {G.~E. S.-T.}\ \bibnamefont {Ana-Sun\v{c}ana~Smith}}, \ and\
  \bibinfo {author} {\bibfnamefont {S.}~\bibnamefont {Torquato}},\ }\href
  {\doibase https://doi.org/10.1038/s41467-019-08360-5} {\bibfield  {journal}
  {\bibinfo  {journal} {Nat. Commun.}\ }\textbf {\bibinfo {volume} {10}},\
  \bibinfo {pages} {811} (\bibinfo {year} {2019})}\BibitemShut {NoStop}%
\bibitem [{\citenamefont {Park}\ and\ \citenamefont
  {Shibutani}(2007)}]{Park2007}%
  \BibitemOpen
  \bibfield  {author} {\bibinfo {author} {\bibfnamefont {J.}~\bibnamefont
  {Park}}\ and\ \bibinfo {author} {\bibfnamefont {Y.}~\bibnamefont
  {Shibutani}},\ }\href {\doibase
  https://doi.org/10.1016/j.intermet.2006.05.005} {\bibfield  {journal}
  {\bibinfo  {journal} {Intermetallics}\ }\textbf {\bibinfo {volume} {15}},\
  \bibinfo {pages} {187} (\bibinfo {year} {2007})}\BibitemShut {NoStop}%
\bibitem [{\citenamefont {Ruocco}\ \emph {et~al.}(1991)\citenamefont {Ruocco},
  \citenamefont {Sampoli},\ and\ \citenamefont {Vallauri}}]{Ruocco1991}%
  \BibitemOpen
  \bibfield  {author} {\bibinfo {author} {\bibfnamefont {G.}~\bibnamefont
  {Ruocco}}, \bibinfo {author} {\bibfnamefont {M.}~\bibnamefont {Sampoli}}, \
  and\ \bibinfo {author} {\bibfnamefont {R.}~\bibnamefont {Vallauri}},\ }\href
  {\doibase https://doi.org/10.1016/0022-2860(91)85033-Y} {\bibfield  {journal}
  {\bibinfo  {journal} {J. Mol. Struct.}\ }\textbf {\bibinfo {volume} {250}},\
  \bibinfo {pages} {259} (\bibinfo {year} {1991})}\BibitemShut {NoStop}%
\bibitem [{\citenamefont {Galindo-Torres}\ and\ \citenamefont
  {Pedroso}(2010)}]{Torres2010}%
  \BibitemOpen
  \bibfield  {author} {\bibinfo {author} {\bibfnamefont {S.~A.}\ \bibnamefont
  {Galindo-Torres}}\ and\ \bibinfo {author} {\bibfnamefont {D.~M.}\
  \bibnamefont {Pedroso}},\ }\href {\doibase 10.1103/PhysRevE.81.061303}
  {\bibfield  {journal} {\bibinfo  {journal} {Phys. Rev. E}\ }\textbf {\bibinfo
  {volume} {81}},\ \bibinfo {pages} {061303} (\bibinfo {year}
  {2010})}\BibitemShut {NoStop}%
\bibitem [{\citenamefont {Derlet}(2020)}]{Derlet2020}%
  \BibitemOpen
  \bibfield  {author} {\bibinfo {author} {\bibfnamefont {P.~M.}\ \bibnamefont
  {Derlet}},\ }\href {\doibase 10.1103/PhysRevMaterials.4.125601} {\bibfield
  {journal} {\bibinfo  {journal} {Phys. Rev. Mater.}\ }\textbf {\bibinfo
  {volume} {4}},\ \bibinfo {pages} {125601} (\bibinfo {year}
  {2020})}\BibitemShut {NoStop}%
\bibitem [{\citenamefont {Dierking}\ \emph {et~al.}(2021)\citenamefont
  {Dierking}, \citenamefont {Flatley},\ and\ \citenamefont
  {Greenhalgh}}]{Ingo2021}%
  \BibitemOpen
  \bibfield  {author} {\bibinfo {author} {\bibfnamefont {I.}~\bibnamefont
  {Dierking}}, \bibinfo {author} {\bibfnamefont {A.}~\bibnamefont {Flatley}}, \
  and\ \bibinfo {author} {\bibfnamefont {D.}~\bibnamefont {Greenhalgh}},\
  }\href {\doibase https://doi.org/10.1016/j.molliq.2021.116553} {\bibfield
  {journal} {\bibinfo  {journal} {Journal of Molecular Liquids}\ }\textbf
  {\bibinfo {volume} {335}},\ \bibinfo {pages} {116553} (\bibinfo {year}
  {2021})}\BibitemShut {NoStop}%
\bibitem [{\citenamefont {Focassio}\ \emph
  {et~al.}(2021{\natexlab{b}})\citenamefont {Focassio}, \citenamefont
  {Schleder}, \citenamefont {Crasto~de Lima}, \citenamefont {Lewenkopf},\ and\
  \citenamefont {Fazzio}}]{Focassio2021b}%
  \BibitemOpen
  \bibfield  {author} {\bibinfo {author} {\bibfnamefont {B.}~\bibnamefont
  {Focassio}}, \bibinfo {author} {\bibfnamefont {G.~R.}\ \bibnamefont
  {Schleder}}, \bibinfo {author} {\bibfnamefont {F.}~\bibnamefont {Crasto~de
  Lima}}, \bibinfo {author} {\bibfnamefont {C.}~\bibnamefont {Lewenkopf}}, \
  and\ \bibinfo {author} {\bibfnamefont {A.}~\bibnamefont {Fazzio}},\ }\href
  {\doibase 10.1103/PhysRevB.104.214206} {\bibfield  {journal} {\bibinfo
  {journal} {Phys. Rev. B}\ }\textbf {\bibinfo {volume} {104}},\ \bibinfo
  {pages} {214206} (\bibinfo {year} {2021}{\natexlab{b}})}\BibitemShut
  {NoStop}%
\bibitem [{\citenamefont {Ebert}\ \emph {et~al.}(2002)\citenamefont {Ebert},
  \citenamefont {Musgrave}, \citenamefont {Peachey}, \citenamefont {Perlin},\
  and\ \citenamefont {Worley}}]{procedural-book}%
  \BibitemOpen
  \bibfield  {author} {\bibinfo {author} {\bibfnamefont {D.~S.}\ \bibnamefont
  {Ebert}}, \bibinfo {author} {\bibfnamefont {F.~K.}\ \bibnamefont {Musgrave}},
  \bibinfo {author} {\bibfnamefont {D.}~\bibnamefont {Peachey}}, \bibinfo
  {author} {\bibfnamefont {K.}~\bibnamefont {Perlin}}, \ and\ \bibinfo {author}
  {\bibfnamefont {S.}~\bibnamefont {Worley}},\ }\href@noop {} {\emph {\bibinfo
  {title} {Texturing and Modeling: A Procedural Approach}}},\ \bibinfo
  {edition} {3rd}\ ed.\ (\bibinfo  {publisher} {Morgan Kaufmann Publishers
  Inc.},\ \bibinfo {address} {San Francisco, CA, USA},\ \bibinfo {year}
  {2002})\BibitemShut {NoStop}%
\bibitem [{\citenamefont {Perlin}(1985)}]{Perlin1985}%
  \BibitemOpen
  \bibfield  {author} {\bibinfo {author} {\bibfnamefont {K.}~\bibnamefont
  {Perlin}},\ }\href {\doibase https://doi.org/10.1145/325165.325247}
  {\bibfield  {journal} {\bibinfo  {journal} {ACM SIGGRAPH Computer Graphics}\
  }\textbf {\bibinfo {volume} {19}},\ \bibinfo {pages} {287} (\bibinfo {year}
  {1985})}\BibitemShut {NoStop}%
\bibitem [{\citenamefont {Perlin}(2002)}]{improved-perlin}%
  \BibitemOpen
  \bibfield  {author} {\bibinfo {author} {\bibfnamefont {K.}~\bibnamefont
  {Perlin}},\ }\href {\doibase 10.1145/566654.566636} {\bibfield  {journal}
  {\bibinfo  {journal} {ACM Trans. Graph.}\ }\textbf {\bibinfo {volume} {21}},\
  \bibinfo {pages} {681–682} (\bibinfo {year} {2002})}\BibitemShut {NoStop}%
\bibitem [{\citenamefont {Loring}\ and\ \citenamefont
  {Hastings}(2011)}]{Loring_2010}%
  \BibitemOpen
  \bibfield  {author} {\bibinfo {author} {\bibfnamefont {T.~A.}\ \bibnamefont
  {Loring}}\ and\ \bibinfo {author} {\bibfnamefont {M.~B.}\ \bibnamefont
  {Hastings}},\ }\href {\doibase 10.1209/0295-5075/92/67004} {\bibfield
  {journal} {\bibinfo  {journal} {EPL}\ }\textbf {\bibinfo {volume} {92}},\
  \bibinfo {pages} {67004} (\bibinfo {year} {2011})}\BibitemShut {NoStop}%
\bibitem [{Sup()}]{SupMat}%
  \BibitemOpen
  \href@noop {} {}\bibinfo {note} {See Supplemental Material at [URL will be
  inserted by publisher] for additional information on the implementation of
  the Perlin Noise, topological aspects of the Agarwalla-Shenoy model,
  evolution of band parameters with amorphization, and on the nature of the
  topological states.}\BibitemShut {Stop}%
\bibitem [{\citenamefont {Harrison}(1989)}]{harrison1989electronic}%
  \BibitemOpen
  \bibfield  {author} {\bibinfo {author} {\bibfnamefont {W.}~\bibnamefont
  {Harrison}},\ }\href
  {https://books.google.com.br/books/about/Electronic_Structure_and_the_Properties.html?id=orAPAQAAMAAJ&redir_esc=y}
  {\emph {\bibinfo {title} {Electronic Structure and the Properties of Solids:
  The Physics of the Chemical Bond}}}\ (\bibinfo  {publisher} {Dover
  Publications},\ \bibinfo {year} {1989})\BibitemShut {NoStop}%
\end{thebibliography}%


%apsrev4-2.bst 2019-01-14 (MD) hand-edited version of apsrev4-1.bst
%Control: key (0)
%Control: author (8) initials jnrlst
%Control: editor formatted (1) identically to author
%Control: production of article title (0) allowed
%Control: page (0) single
%Control: year (1) truncated
%Control: production of eprint (0) enabled
\begin{thebibliography}{3}%
\makeatletter
\providecommand \@ifxundefined [1]{%
 \@ifx{#1\undefined}
}%
\providecommand \@ifnum [1]{%
 \ifnum #1\expandafter \@firstoftwo
 \else \expandafter \@secondoftwo
 \fi
}%
\providecommand \@ifx [1]{%
 \ifx #1\expandafter \@firstoftwo
 \else \expandafter \@secondoftwo
 \fi
}%
\providecommand \natexlab [1]{#1}%
\providecommand \enquote  [1]{``#1''}%
\providecommand \bibnamefont  [1]{#1}%
\providecommand \bibfnamefont [1]{#1}%
\providecommand \citenamefont [1]{#1}%
\providecommand \href@noop [0]{\@secondoftwo}%
\providecommand \href [0]{\begingroup \@sanitize@url \@href}%
\providecommand \@href[1]{\@@startlink{#1}\@@href}%
\providecommand \@@href[1]{\endgroup#1\@@endlink}%
\providecommand \@sanitize@url [0]{\catcode `\\12\catcode `\$12\catcode
  `\&12\catcode `\#12\catcode `\^12\catcode `\_12\catcode `\%12\relax}%
\providecommand \@@startlink[1]{}%
\providecommand \@@endlink[0]{}%
\providecommand \url  [0]{\begingroup\@sanitize@url \@url }%
\providecommand \@url [1]{\endgroup\@href {#1}{\urlprefix }}%
\providecommand \urlprefix  [0]{URL }%
\providecommand \Eprint [0]{\href }%
\providecommand \doibase [0]{https://doi.org/}%
\providecommand \selectlanguage [0]{\@gobble}%
\providecommand \bibinfo  [0]{\@secondoftwo}%
\providecommand \bibfield  [0]{\@secondoftwo}%
\providecommand \translation [1]{[#1]}%
\providecommand \BibitemOpen [0]{}%
\providecommand \bibitemStop [0]{}%
\providecommand \bibitemNoStop [0]{.\EOS\space}%
\providecommand \EOS [0]{\spacefactor3000\relax}%
\providecommand \BibitemShut  [1]{\csname bibitem#1\endcsname}%
\let\auto@bib@innerbib\@empty
%</preamble>
\bibitem [{\citenamefont {Perlin}(1985)}]{Perlin1985}%
  \BibitemOpen
  \bibfield  {author} {\bibinfo {author} {\bibfnamefont {K.}~\bibnamefont
  {Perlin}},\ }\bibfield  {title} {\bibinfo {title} {An image synthesizer},\
  }\href {https://doi.org/https://doi.org/10.1145/325165.325247} {\bibfield
  {journal} {\bibinfo  {journal} {ACM SIGGRAPH Computer Graphics}\ }\textbf
  {\bibinfo {volume} {19}},\ \bibinfo {pages} {287} (\bibinfo {year}
  {1985})}\BibitemShut {NoStop}%
\bibitem [{\citenamefont {Agarwala}\ and\ \citenamefont
  {Shenoy}(2017)}]{Agarwala2017}%
  \BibitemOpen
  \bibfield  {author} {\bibinfo {author} {\bibfnamefont {A.}~\bibnamefont
  {Agarwala}}\ and\ \bibinfo {author} {\bibfnamefont {V.~B.}\ \bibnamefont
  {Shenoy}},\ }\bibfield  {title} {\bibinfo {title} {Topological insulators in
  amorphous systems},\ }\href {https://doi.org/10.1103/PhysRevLett.118.236402}
  {\bibfield  {journal} {\bibinfo  {journal} {Phys. Rev. Lett.}\ }\textbf
  {\bibinfo {volume} {118}},\ \bibinfo {pages} {236402} (\bibinfo {year}
  {2017})}\BibitemShut {NoStop}%
\bibitem [{\citenamefont {Harrison}(1989)}]{harrison1989electronic}%
  \BibitemOpen
  \bibfield  {author} {\bibinfo {author} {\bibfnamefont {W.}~\bibnamefont
  {Harrison}},\ }\href
  {https://books.google.com.br/books/about/Electronic_Structure_and_the_Properties.html?id=orAPAQAAMAAJ&redir_esc=y}
  {\emph {\bibinfo {title} {Electronic Structure and the Properties of Solids:
  The Physics of the Chemical Bond}}}\ (\bibinfo  {publisher} {Dover
  Publications},\ \bibinfo {year} {1989})\BibitemShut {NoStop}%
\end{thebibliography}%
%%%%%%%%%%%%%%%%%%%%%%%%%%%%%%%%%%%%%%%%%%%%%%%%%%%%%%%%%

\end{document}

\title{Supplemental Material to 
\texorpdfstring{\\}{}
``Structure-driven phase transitions in noncrystalline topological insulators"}

\author{Victor Regis}
\affiliation{Jožef Stefan Institute, 1000 Ljubljana, Slovenia}

\author{Victor Velasco}
\affiliation{Instituto de F\'isica, Universidade Federal do Rio de Janeiro, Caixa Postal 68528, 21941-972 Rio de Janeiro, Brazil}
\affiliation{School of Pharmacy, Physics Unit, Università di Camerino, Via Madonna delle Carceri 9, 62032 Camerino, Italy}

\author{Marcello B. Silva Neto}
\affiliation{Instituto de F\'isica, Universidade Federal do Rio de Janeiro, Caixa Postal 68528, 21941-972 Rio de Janeiro, Brazil}
\affiliation{Laboratório Nacional de Nanotecnologia, CNPEM, 13083-100, Campinas, São Paulo, Brazil}

\author{Caio Lewenkopf}
\affiliation{Instituto de F\'isica, Universidade Federal Fluminense, 24210-346 Niter\'oi, Brazil}

\date{\today}

\maketitle

%%%%%%%%%%%%%%%%%%%%%%%%%%%%%%%%%%%%%%%%%%%%%%%%%%%%%%%%%%%%%%%%%%%%%%%%%%%
\section{The Perlin noise}
\label{SM: Perlin}

One of the simplest ways of sampling random numbers is to consider a uniform distribution. A set of uncorrelated points drawn from this kind of distribution in two-dimensions looks like a shapeless, locally discontinuous noise. For the purposes of the present work, uncorrelated noise is not useful to continuously deform a crystalline lattice to a random system, passing through the paracrystalline phases, since the correlation between sites quickly vanishes. 
To properly perform the transitions one needs to use a correlated noise map.
We use the Perlin noise \cite{Perlin1985}. 
This algorithm, that is very well-known in computer graphics, is extremely powerful in generating textures in a procedural manner, meaning without them being manually made by an artist or designer, and is often used in procedural content generation, like in games and movies. 
As a measure of the impact of the algorithm, it is worth to mention that the Ken Perlin won the Technical Achievement Award from the Academy of Motion Picture Arts and Sciences in 1997 for his work on procedural texture. Here, we explain in details the application of the Perlin noise for the continuous deformation of a crystalline lattice towards a random system.

The algorithm can be generated in any number of dimensions. 
Here, we are interested in the two dimensional case. As an input, the PN gets a set of floating points depending on the number of dimensions and returns a value in a certain range that is usually defined to be between $-1.0$ and $+1.0$. In two dimensions, the inputs are the $x$ and $y$ positions of the points in the crystal.  As stated in the main text, the algorithm can be divided in two parts: the randomization and the amorphization procedures. The first part is responsible for the generation of the noise map following the Perlin noise algorithm and the amorphization is the usage of the noise to displace the atoms on the lattice accordingly. Therefore, in order to generate a texture, we need to iterate through every lattice point in the crystal calling the Perlin noise map for each one and let the points move around.

\begin{figure}
\centering
\includegraphics[width=0.95\columnwidth]{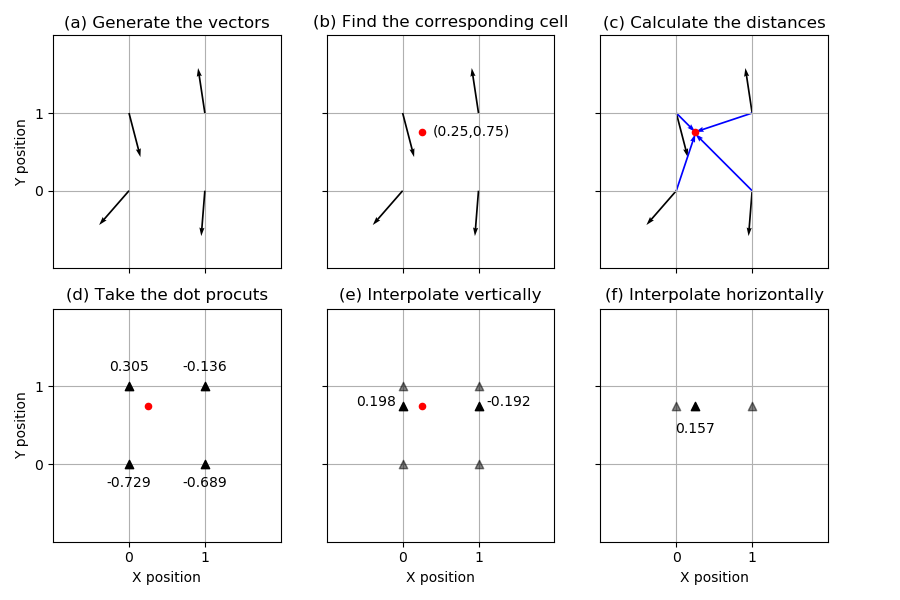}
\caption{(a) The four vectors are generated at the corners of the cell. (b) The point is calculated to belong within the cell. (c) The distance vector is calculated from the corners to the input point, indicated as blue arrows. (d) The dot product between the distance vectors and their corresponding corners is calculated. (e) Interpolate vertically the pairs on the left and right. The original dot products are indicated as lighters triangles and the interpolated results are the solid triangles. (f) Interpolate the resulting pair horizontally to obtain the noise at the point. The new result is shown as a solid triangle.}
\label{fig:perlin-proc}
\end{figure}

For the randomization procedure, we start with a $N \times N$ square grid and divide it into $N^2$ square cells of side $a = 1$. For each cell, we generate four unitary random vectors, $\mathbf{v_{ij}}$, on its corners, as in Fig. \ref{fig:perlin-proc}a). For an input point $\mathbf{p} = (x,y)$, identify to which cell it belongs to by calculating the floor and ceiling values of its coordinates. Next, calculate the vectors that define the distances from the corners to the input point, $\mathbf{c}_{ij}$, and calculate their dot product with their respective unitary vector, namely, $\delta_{ij} = \mathbf{v}_{ij} \cdot \mathbf{c}_{ij}$, as indicated in Fig.\ref{fig:perlin-proc}(c--d). Here $i$ and $j$ are auxiliary indices to identify the corners, namely $v_{0,0}$ in the bottom-left, $v_{1,0}$ in the top-left, $v_{0,1}$ in the bottom-right and $v_{1,1}$ in the top-right corners. At this stage, there are four dot products associated with the input point $\mathbf{p}$. These values must be reduced to a single noise value. First, we interpolate the left-side values weighted according to their $y$-coordinate relative to $\mathbf{p}$, then we do the same for the right-hand side pair. Finally, we interpolate the left and the right, weighted according to their $x$-position, as schematically shown in Fig.\ref{fig:perlin-proc}(e--f), ending with a single value for the noise of the representative point.

\begin{figure}
\centering
\includegraphics[width=0.95\columnwidth]{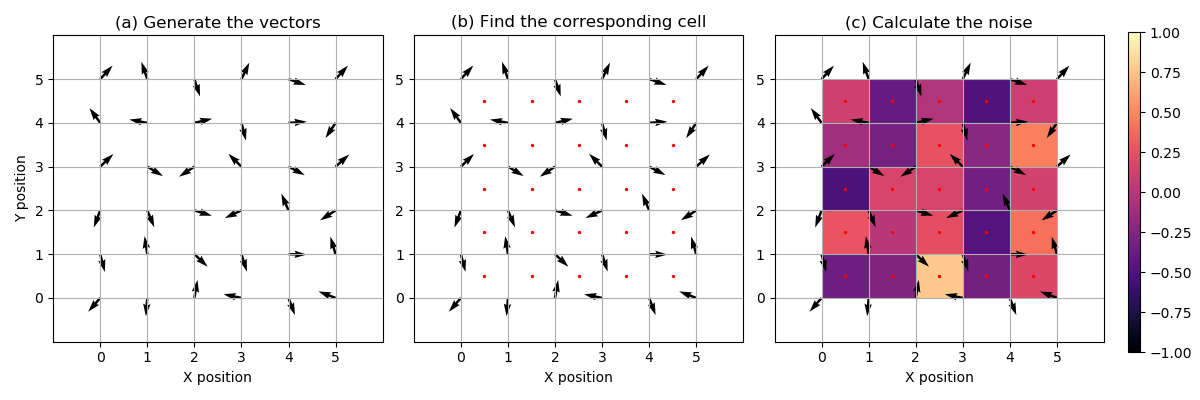}
\caption{An example of the Perlin Noise for a $5\times5$ set of points. (a) The vectors are generated at the corners of each cell. (b) The points are scattered throughout the space and their corresponding cell is identified. In this case, there is only one point per cell and each point is located in the middle of their cell. (c) Using the Perlin noise procedure, the noise values are calculated for each input point, the values are indicated by the color bar. Notice that neighboring cells are of similar color (value) and that cells with vectors pointing inwards have high position values and outwards, negative values.}
\label{fig: Perlin_full}
\end{figure}

The weight function is called the {\it smoothstep} function and it is chosen so that the further the point is from a corner, the less impactful its unitary vector is on the resulting noise value. It is defined as $S(t) = 3t^2-2t^3$ with $t$ being a coordinate of the vector $\boldsymbol{p} - c_{0,0}$, where $c_{0,0}$ is the position of the corner on the lower left side of the cell. The variable $t$ is the $x$ coordinate when the interpolation is horizontal and the $y$ coordinate when is vertical. Since the cells are squares of unitary sides, $t\in[0,1]$, it is zero when the point is at the lower (leftmost) edge of the cell and unity when the point is at the upper (rightmost) edge. The linear interpolation is given by
$L(s,a,b) = a + s\left(b-a \right)$, where $s$ is the weight value obtained from the smoothstep function $S(t)$. In the vertical case, $a$ is the dot value at the lower corner and $b$ is the dot value at the upper corner in the pair. In the horizontal case, we use $a$ as the leftmost value and $b$ as the rightmost value. The weighted linear interpolation ensures a smooth transition between noise values from a cell to a neighboring one. After the dot products and the linear interpolations, the output is a noise value, $n(x,y)$ at the position of the input point. An example for a complete generation of the noise map following the procedure described above for a $5\times 5$ set of points in 2D in shown on Fig. \ref{fig: Perlin_full}.

It is worth mentioning that by following the PN procedure, the further apart two points are, the less correlated they are. This comes from the fact that the corner vectors generated at the first step of the procedure are fixed for a given generation of a noise map $n(x,y)$. Thus, if two points do not share a common corner vector, for instance points that are more that two cells apart, the correlation among these points is weaker than between points that are within the same or neighboring cells. In order to map all points of a given set into correlated noise values, the set has to be mapped into a single cell, so that all noise will be calculated using the same
corner vectors. This can be done by multiplying the set by a constant factor $\alpha$, usually set by the inverse of the largest coordinate value. Thus, the multiplicative factor $\alpha$ controls the correlation length. It is important to notice that the greater the factor $\alpha$ is, less correlated is the noise map.

There are a couple of features of the method worth being emphasized. The first is a controllable correlation length (by multiplying the set of input points by a factor $\alpha$, the correlation length of the output noise can be regulated). The second are the octaves. The noise map we have discussed thus far is a map with a single octave and it suffers from not being as realistic as one might want. For one octave, the noise map \textit{texture} does not look as \textit{noisy} as one would expect. To work around this issue, one should calculate the noise map for $\alpha = 2^{N-1}$, where $N$ is the order of the octave, divide the output noise by $\alpha$ and sum it to the preceding noise map. Adding higher-order octaves improves the \textit{texture} of the map making it look less like a simple combination of trigonometric functions.

\begin{figure}
\centering
\includegraphics[width=0.95\columnwidth]{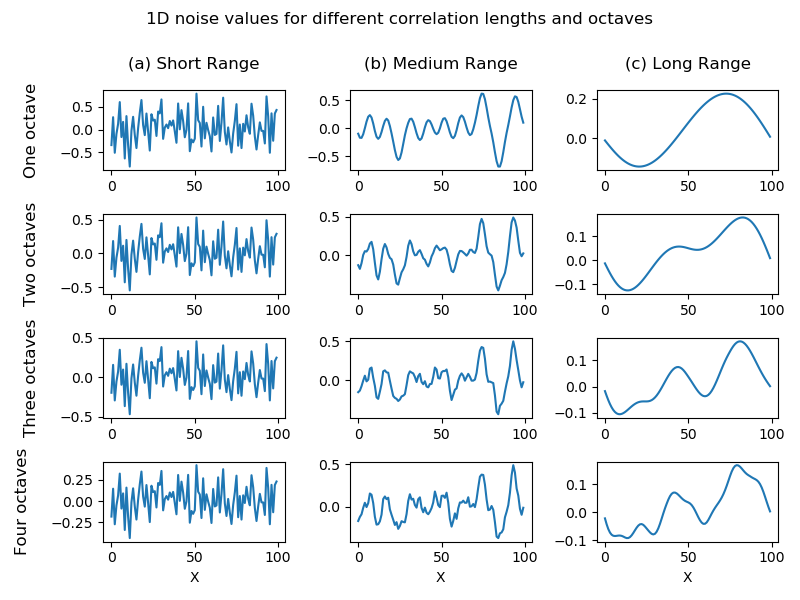}
\caption{Transversal cut along the $y$-axis of a noise map for $100 \times 100$ 2D points spaced out evenly. The vertical axis represents the noise value. (a) Noise map for short range correlation length with $\alpha = 1$ and  $r_{corr} =  1$. (b) For $\alpha = 0.1$ and  $r_{corr} = 10$ and (c) for $\alpha = 0.01$ and the correlation length  $r_{corr} = 100$.}
\label{fig:octaves}
\end{figure}

\begin{figure}[ht]
\centering
\includegraphics[width=0.95\columnwidth]{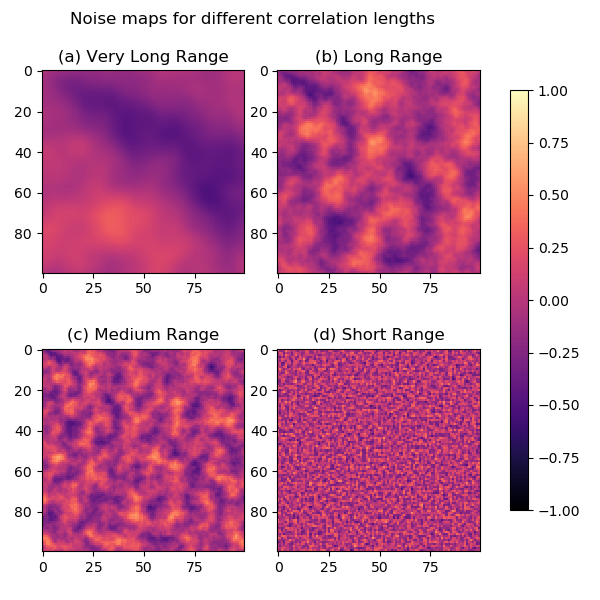}
\caption{Noise maps for a $100 \times 100$ grid of points with four octaves. The colorbar shows the intensity of the noise. (a) The correlation length is set to $r_{corr}= 100$ making the noise map smooth with no large change in value between neighboring sites. (b) By decreasing the length to $r_{corr} = 20$, the map gains more valleys and peaks. (c) When the length is $r_{corr} \approx 10$, the map has several valleys and peaks nearby, however, it is still locally correlated. (d) For $r_{corr} = 2$, the noise is uncorrelated enough that is looks like white noise, i.e., noise from a uniform distribution. }
\label{fig:maps}
\end{figure}

In Fig. \ref{fig:octaves} we show the representative 1D case for a $100 \times 100$ 2D system, for three different correlation lengths, namely, short (a), medium (b) and long (c) correlation lengths and four different octaves added to the noise maps. In Fig. \ref{fig:octaves}a), we set the short correlation length via $\alpha = 1$, which gives a radius of correlation $r_{corr} = 1$, which counts the number of atoms per cell. Since in this case each site is located at an individual cell and their noise values are weakly correlated, the noise map is very similar to a uniform  distribution map. In this case, adding octaves has no significant effect on the final noise map. For $\alpha = 0.1$ and $r_{corr} = 10$ in Fig. \ref{fig:octaves}b), there are ten sites per cell, hence the noise is locally correlated and there is no discontinuous change in value for neighboring points. For the one octave case, the noise map looks like a sum of trigonometric functions, however by adding more octaves we add more complexity to it, making it less predictable. Finally, in Fig. \ref{fig:octaves}c) $\alpha = 0.01$ and the correlation length is $r_{corr} = 100$, making the noise highly correlated. The change in value from site to  site is very small and gradual. For the one octave case, the noise is  very predictable and looks like a sine function. As octaves are added, it becomes \textit{noisier} and more realistic from the point of view of the procedural generation algorithms. As another representative case, we show in Fig. \ref{fig:maps} a set of 2D noise maps with diﬀerent correlation lengths.

Finally, we describe the amorphization procedure that produces the deformation of the crystal lattice. We divided this part of the procedure in $t = 100$ iteration steps. It is worth mentioning that after each iteration, the correlation length of the system is decreased. At each iteration, a noise value is drawn from the map and summed to the corresponding site position. For instance, the $i-$th site with coordinates $(x_i, y_i)$ has a corresponding noise of value $n(x_i, y_i, t)$ draw from the map at the given iteration. Since at each iteration the correlation between neighboring sites is decreased, the system slowly moves from crystalline to a random one. 

The structural transitions are obtained following the algorithm: at each iteration $t$, a site at a position $\mathbf{p}_{t-1}$ at the previous iteration is selected. From the noise map $n(x,y,t)$, the corresponding offset value, $\mathbf{\delta}_p$, is drawn. The tentative new site position is $\mathbf{p}_t = \mathbf{p}_{t-1} + \mathbf{\delta}_p$, since the algorithm checks if there are any other sites within a circle centered at the new position with radius $0.9a_0$, where $a_0$ is the is the lattice parameter of the crystal structure, set to unity in this study. If there are no sites, then $\mathbf{p}_t$ is an accepted new position. If there are any other sites in the circle, then $\mathbf{p}_t = \mathbf{p}_{t-1} + 0.1\mathbf{\delta}_p$, that is only a small fraction of the offset is added to the previous position. To reach the amorphous phase we must ensure that short-range correlations endure throughout the process and that the long-range ones are eliminated. Hence, the ﬁrst condition ensures that the former is true, while the adding of a small oﬀset despite new neighbors within the circle assures the latter. Finally, to reach the random configuration, at an arbitrary number of iteration, in our case half of the total number of iterations $t$, we linearly increase the randomization weight and let sites be offset independently of their neighbors. 

%%%%%%%%%%%%%%%%%%%%%%%%%%%%%%%%%%%%%%%%%%%%%%%%%%%%%%%%%%%%%%%%%%%%%%%%
\section{The Agarwala-Shenoy Model}

Let us discuss the topological aspects associated to the Agarwala-Shenoy model \cite{Agarwala2017}. 
The model is defined by the two-dimensional tight-binding Hamiltonian introduced in Eq.~(1) of the main text. Before proceeding, some comments are in order: in Ref. \cite{Agarwala2017} the hopping $t_1=1$, but we shall keep the notation $t_1$ for the sake of clarity and dimensionality. 
Also, in the original formulation, the separation between two sites is characterized by a radius $|{\bf r}_{IJ}|=r$ and an angle $\theta$ which are continuous variables, $0<r<R$ and $0<\theta<2\pi$. Here, instead, we shall consider a pristine square lattice, such that $r=a$, the lattice spacing, and $\theta=0,\pi/2,\pi,3\pi/2$, for the nearest neighbors. Finally, for the prefactor, $t(r)$, we shall consider it $t(r)=1$ and the final Hamiltonian will be fully determined from $T_{\alpha\beta}(\theta)$.

The main purpose of this appendix is to obtain the reciprocal space Hamiltonian
\begin{equation}
{\cal H}=\sum_{{\bf k}}\sum_{\alpha\beta}H_{\alpha\beta}({\bf k})c^\dagger_{{\bf k},\alpha}c_{{\bf k},\beta},
\end{equation}
for the case of nearest neighbors, $\sum_{I,J}\rightarrow \sum_{\langle I,J\rangle}$.
Since this is a $2\times 2$ problem, the Hamiltonian $H_{\alpha\beta}({\bf k})$ can be decomposed in terms of the Pauli matrices, $\sigma_{\alpha\beta}^\mu$ with $\mu=0,x,y,z$, in the form
\begin{equation}
    H_{\alpha\beta}({\bf k})=d_\mu({\bf k})\sigma_{\alpha\beta}^\mu,
\end{equation}
where we adopt the matrix forms below
\begin{widetext}
\begin{equation}
    \sigma_{\alpha\beta}^0=
    \begin{pmatrix}
        1 & 0 \\
        0 & 1
    \end{pmatrix},\quad
    \sigma_{\alpha\beta}^x=
    \begin{pmatrix}
        0 & 1 \\
        1 & 0
    \end{pmatrix},\quad
    \sigma_{\alpha\beta}^y=
    \begin{pmatrix}
        0 & -i \\
        i & 0
    \end{pmatrix},\quad
    \sigma_{\alpha\beta}^z=
    \begin{pmatrix}
        1 & 0 \\
        0 & -1
    \end{pmatrix}.
\end{equation}
\end{widetext}
By looking at the hopping matrix $T_{\alpha\beta}(\theta)$, some considerations can be made for the components $d_\mu({\bf k})$, namely:

\begin{itemize}
    \item The on-site matrix can be immediately written as $\epsilon_{\alpha\beta}=d_\mu\sigma_{\alpha\beta}^\mu$, where the vector $d_\mu$ does not have any ${\bf k}-$dependence (since $I=J$) and all of its coefficients are real $d_\mu=(0,\lambda,\lambda,2+M)$;

    \item The spin-independent, intra-orbital hopping $t_2$ is ${\bf k}-$dependent (since $I\neq J$) and contribute to the diagonal components, $d_0({\bf k})$ (since it is independent of the spin projection $\alpha=\beta$). Furthermore since it is purely real, its ${\bf k}-$dependence is in terms of $\cos{k_x}$ and $\cos{k_y}$; 

    \item The spin-dependent, intra-orbital hopping $t_1$ is also ${\bf k}-$dependent (since $I\neq J$) and contribute to the $z-$component, $d_z({\bf k})$ (since the hopping has different sign for the two spin components $\alpha,\beta$). Furthermore since it is purely real, its ${\bf k}-$dependece is in terms of $\cos{k_x}$ and $\cos{k_y}$; 

    \item The spin-flip hopping $-i\, e^{i\theta}$ is also ${\bf k}-$dependent (since $I\neq J$) and contribute to the $x,y-$components, $d_{x,y}({\bf k})$ (since $\alpha\neq\beta$). Furthermore because of the $i$ prefactor, its ${\bf k}-$dependence is in terms of $\sin{k_x}$ and $\sin{k_y}$; 

    \item The inter-orbital mixing proportional to $\lambda$ is also ${\bf k}-$dependent (since $I\neq J$) and also contribute to the $x,y-$components, $d_{x,y}({\bf k})$ (since $\alpha\neq\beta$). Furthermore because it is purely real its ${\bf k}-$dependence will be in terms of $\cos{k_x}$ and $\cos{k_y}$; 

\end{itemize}

Let us now calculate explicitly all components of $d_\mu({\bf k})$. We shall use the following prescriptions for Fourier transforms, definition of $\delta-$functions and sum over indices for an $N\times N$ square lattice and coordination $z$:
\begin{eqnarray}
    c^\dagger_{I,\alpha}&=&\frac{1}{\sqrt{N}}\sum_{{\bf k}}e^{i{\bf k}\cdot{\bf r}_I}\; c^\dagger_{{\bf k},\alpha},\nonumber\\
    c_{J,\alpha}&=&\frac{1}{\sqrt{N}}\sum_{{\bf k}^\prime}e^{-i{\bf k}^\prime\cdot{\bf r}_J}\; c_{{\bf k}^\prime,\alpha},\nonumber\\
    \delta_{{\bf k},{\bf k}^\prime}&=&\frac{1}{N}\sum_I e^{i({\bf k}-{\bf k}^\prime)\cdot{\bf r}_I},\nonumber\\
    z&=&\sum_J\delta_{I,J}\quad\mbox{for nearest neighbors}.
\end{eqnarray}
Using the above prescriptions we can now calculate explicitly all components of $d_\mu({\bf k})$.

\begin{enumerate}
    \item The on-site matrix can be immediately written as 
    \begin{eqnarray}
    {\cal H}_{on}&=&\sum_{I}\sum_{\alpha,\beta}
        \epsilon_{\alpha\beta} c^\dagger_{I,\alpha}
        c_{I,\beta}\nonumber\\  
    &=&\sum_{{\bf k},{\bf k}^\prime}\sum_{\alpha,\beta}
    \epsilon_{\alpha\beta}c^\dagger_{{\bf k},\alpha}c_{{\bf k}^\prime,\beta}
    \times\underbrace{\frac{1}{N}\sum_{I}e^{i({\bf k}-{\bf k}^\prime)\cdot{\bf r}_I}}_{\delta_{{\bf k},{\bf k}^\prime}}
    \nonumber\\
    &=&\sum_{{\bf k}}\sum_{\alpha\beta}
    d_\mu\sigma^\mu_{\alpha\beta}c^\dagger_{{\bf k},\alpha}c_{{\bf k},\beta},    
    \end{eqnarray} 
    with $d_\mu=(0,\lambda,\lambda,2+M)$;

    \item The spin-independent $(\alpha=\beta)$, intra-orbital hopping can be written as
    \begin{eqnarray}
        {\cal H}_{0}&=&\frac{1}{2}\sum_{I,J}\sum_{\alpha}t_2
        c^\dagger_{I,\alpha} c_{J,\alpha}\nonumber\\
        &=&\frac{1}{2}\sum_{{\bf k},{\bf k}^\prime}\sum_{\alpha}
        c^\dagger_{{\bf k},\alpha}c_{{\bf k}^\prime,\alpha}
        \left[t_2\gamma({\bf k}^\prime)\right]\times\underbrace{\frac{1}{N}\sum_{I}e^{i({\bf k}-{\bf k}^\prime)\cdot{\bf r}_I}}_{\delta_{{\bf k},{\bf k}^\prime}}\nonumber\\
        &=&\sum_{{\bf k}}\sum_{\alpha,\beta}
        d_0({\bf k})\sigma^0_{\alpha\beta}c^\dagger_{{\bf k},\alpha}c_{{\bf k},\beta}, 
    \end{eqnarray}
    with $d_0({\bf k})=(t_2/2)\gamma({\bf k})$ and
    \begin{eqnarray}
        \gamma({\bf k})&=&\sum_J e^{-i{\bf k}\cdot{\bf r}_J}
        =e^{-i kx a}+e^{+i kx a}+e^{-i ky a}+e^{+i ky a}\nonumber\\
        &=&2(\cos{(k_x a)}+\cos{(k_y a)}),
    \end{eqnarray}
    for nearest neighbors on a square lattice $|{\bf r}_{IJ}|=a$.

    \item The spin-dependent, intra-orbital hopping $t_1$ can be written as
    \begin{eqnarray}
        {\cal H}_{z}&=&-\frac{1}{2}\sum_{I,J}\sum_{\alpha,\beta}t_1\sigma^z_{\alpha\beta}
        c^\dagger_{I,\alpha} c_{J,\beta}\nonumber\\
        &=&-\frac{1}{2}\sum_{{\bf k},{\bf k}^\prime}\sum_{\alpha,\beta}t_1\sigma^z_{\alpha\beta}
        c^\dagger_{{\bf k},\alpha}c_{{\bf k}^\prime,\beta}
        \gamma({\bf k}^\prime)\times\underbrace{\frac{1}{N}\sum_{I}e^{i({\bf k}-{\bf k}^\prime)\cdot{\bf r}_I}}_{\delta_{{\bf k},{\bf k}^\prime}}\nonumber\\
        &=&\sum_{{\bf k}}\sum_{\alpha,\beta}
        d_z({\bf k})\sigma^z_{\alpha\beta}c^\dagger_{{\bf k},\alpha}c_{{\bf k},\beta}, 
    \end{eqnarray}
    with $d_z({\bf k})=-(t_1/2)\gamma({\bf k})$ and $\gamma({\bf k})$ is the same as calculated above.

    \item The spin-flip hopping with $-i\, e^{i\theta_{x,y}^{\alpha\beta}}$, such that $\theta_{x,y}^{\alpha\beta}=-\theta_{x,y}^{\beta\alpha}$ can be written as
    \begin{eqnarray}
        {\cal H}_{x,y}&=&-\frac{1}{2}\sum_{I,J}\sum_{\alpha,\beta}i\,e^{i\theta_{x,y}^{\alpha\beta}}
        c^\dagger_{I,\alpha} c_{J,\beta}\nonumber\\
        &=&-\frac{1}{2}\sum_{{\bf k},{\bf k}^\prime}\sum_{\alpha,\beta}
        c^\dagger_{{\bf k},\alpha}c_{{\bf k}^\prime,\beta}
        \sum_J i e^{i\theta_{x,y}^{\alpha\beta}}\,e^{-i{\bf k}\cdot{\bf r}_J}\nonumber\\
        &\times&\underbrace{\frac{1}{N}\sum_{I}e^{i({\bf k}-{\bf k}^\prime)\cdot{\bf r}_I}}_{\delta_{{\bf k},{\bf k}^\prime}}\nonumber\\
        &=&\sum_{{\bf k}}\sum_{\alpha,\beta}
        d_{x,y}({\bf k})\sigma^{x,y}_{\alpha\beta}c^\dagger_{{\bf k},\alpha}c_{{\bf k},\beta}, 
    \end{eqnarray}
    with $d_{x,y}({\bf k})=-(1/2)\eta_{x,y}({\bf k})$ and 
    \begin{equation}
        \eta_{x,y}({\bf k})=i\sum_J e^{i\theta_{x,y}}\,e^{-i{\bf k}\cdot{\bf r}_J}.
    \end{equation}
    Since $\cos{(\pm\theta)}=\cos{\theta}$ the $\pm$ sign is irrelevant for the coefficient of $\sigma^x_{\alpha\beta}$ (which is purely real) and since $\sin{(\pm\theta)}=\pm\sin{\theta}$ the $\pm$ sign is absorbed in the definition of $\sigma^y_{\alpha\beta}$ (which is purely imaginary). For this reason from now on we shall replace $\theta_{x,y}^{\alpha\beta}$ by $\theta_{x,y}=0,\pi/2,\pi,3\pi/2$.
    
    If we use $\theta_x=0$ for the right neighbor, $\theta_y=\pi/2$ for the neighbor above, $\theta_x=\pi$ for the left neighbor, and $\theta_y=3\pi/2$ for the neighbor below we obtain
    \begin{eqnarray}
        \eta_{x}({\bf k})&=&i\left\{e^{i0}e^{-i k_{x} a}+e^{i\pi}e^{+i k_{x} a}\right\}\nonumber\\
        &=&-\left\{\frac{e^{-i k_{x} a}-e^{+i k_{x} a}}{i}\right\}=2\sin{(k_x a)},
    \end{eqnarray}
    and
    \begin{eqnarray}
        \eta_{y}({\bf k})&=&i\left\{e^{i\pi/2}e^{-i k_{y} a}+e^{i3\pi/2}e^{+i k_{y} a}\right\}\nonumber\\
        &=&-\left\{\frac{e^{-i k_{y} a}-e^{+i k_{y} a}}{i}\right\}=2\sin{(k_y a)}.
    \end{eqnarray}

    \item The inter-orbital mixing contains the term $\lambda\left[\sin^2\theta(1\pm i)-1\right]$ where the $\pm$ signs are relative to the $\alpha\beta$ (or $+$) and $\beta\alpha$ (or $-$) matrix elements. The $\alpha\beta$ (or $+$) part can be written as
    \begin{eqnarray}
        {\cal H}^\lambda_{x,y}&=&\frac{1}{2}\sum_{I,J}\sum_{\alpha,\beta}
        \lambda\left[\sin^2\theta_{x,y}(1+i)-1\right]
        c^\dagger_{I,\alpha} c_{J,\beta}\nonumber\\
        &=&\frac{1}{2}\sum_{{\bf k},{\bf k}^\prime}\sum_{\alpha,\beta}
        \lambda\left[\sin^2\theta_{x,y}(1+i)-1\right]
        c^\dagger_{{\bf k},\alpha}c_{{\bf k}^\prime,\beta}\nonumber\\
        &\times&\underbrace{\frac{1}{N}\sum_{I}e^{i({\bf k}-{\bf k}^\prime)\cdot{\bf r}_I}}_{\delta_{{\bf k},{\bf k}^\prime}}\sum_J e^{-i{\bf k}\cdot{\bf r}_J}\nonumber\\
        &=&\sum_{{\bf k}}\sum_{\alpha,\beta}
        d^\lambda_{x,y}({\bf k})\sigma^{x,y}_{\alpha\beta}c^\dagger_{{\bf k},\alpha}c_{{\bf k},\beta}, \nonumber
    \end{eqnarray}
    with $d^\lambda_{x,y}({\bf k})=(1/2)\zeta_{x,y}({\bf k})$ and 
    \begin{equation}
        \zeta_{x,y}({\bf k})=\sum_J (+i)^{\sin^2\theta_{x,y}}\lambda\left[\sin^2\theta_{x,y}(1+i)-1\right]e^{-i{\bf k}\cdot{\bf r}_J}
    \end{equation}
    Again $\theta_x=0$ for the right neighbor, $\theta_y=\pi/2$ for the neighbor above, $\theta_x=\pi$ for the left neighbor, and $\theta_y=3\pi/2$ for the neighbor below such that
    \begin{eqnarray}
        \zeta_{x}({\bf k})&=&(+i)^0\lambda\left[\cancelto{0}{\sin^2(0)}(1+i)-1\right]e^{-i k_x a}\nonumber\\
        &+&(+i)^0\lambda\left[\cancelto{0}{\sin^2(\pi)}(1+i)-1\right]e^{+i k_x a}\nonumber\\
        &=&-\lambda(e^{-i k_x a}+e^{+i k_x a})=-2\lambda\cos{(k_x a)},
    \end{eqnarray}
    and
    \begin{eqnarray}
        \zeta_{y}({\bf k})&=&(+i)^1\lambda\left[\cancelto{1}{\sin^2(\pi/2)}(1+i)-1\right]e^{-i k_y a}\nonumber\\
        &+&(+i)^1\lambda\left[\cancelto{1}{\sin^2(3\pi/2)}(1+i)-1\right]e^{+i k_y a}\nonumber\\
        &=&i\lambda(i e^{-i k_y a}+i e^{+i k_y a})=-2\lambda\cos{(k_y a)}.
    \end{eqnarray}
    The $\beta\alpha$ (or $-$) part can be obtained by replacing $(+i)^{\sin^2\theta_{x,y}}$ for $(-i)^{\sin^2\theta_{x,y}}$ in the expression of $\zeta_{x,y}({\bf k})$.
\end{enumerate}

We can now group all the terms together and write the reciprocal space Hamiltonian
\begin{equation}
{\cal H}=\sum_{{\bf k}}\sum_{\alpha\beta}H_{\alpha\beta}({\bf k})c^\dagger_{{\bf k},\alpha}c_{{\bf k},\beta},
\end{equation}
in terms of the matrix
\begin{equation}
    H({\bf k})=d_0({\bf k}){\cal I}+{\bf d}({\bf k})\cdot\vec{\sigma},
\end{equation}
where
\begin{equation}
    d_0({\bf k})=t_2\left\{\cos{(k_x a)}+\cos{(k_y a)}\right\},
\end{equation}
${\cal I}$ is the identity matrix, $\vec{\sigma}=(\sigma_x,\sigma_y,\sigma_z)$,
and the components of ${\bf d}({\bf k})=(d_x({\bf k}),d_y({\bf k}),d_z({\bf k}))$ are
\begin{eqnarray}
    d_x({\bf k})&=&\lambda(1-\cos{(k_x a)})-\sin{(k_x a)}\nonumber\\
    d_y({\bf k})&=&\lambda(1-\cos{(k_y a)})-\sin{(k_y a)}\nonumber\\
    d_z({\bf k})&=&(2+M)-t_1\left\{\cos{(k_x a)}+\cos{(k_y a)}\right\}.\nonumber\\
\end{eqnarray}
From the above form of the Hamiltonian it is straightforward to obtain the energy spectrum, which is given by two bands corresponding to the expression
\begin{equation}
    E({\bf k})=d_0({\bf k})\pm\sqrt{d^2_x({\bf k})+d^2_y({\bf k})+d^2_z({\bf k})}.
\end{equation}

Regarding the topological structure of the Agarwala-Shenoy model we conclude that:

\begin{itemize}
    \item The Berry curvature associated to such Hamiltonian can be straightforwardly calculated as
    \begin{equation}
        \Omega_{\pm}({\bf k})=\pm\frac{1}{2}\frac{{\bf d}({\bf k})}{|{\bf d}({\bf k})|^3},
    \end{equation}
    and corresponds to a charge $Q=+1$ monopole configuration, located at the origin in reciprocal space (for $Q=-1$ one would have $\Omega_{\pm}({\bf k})\rightarrow\Omega_{\mp}({\bf k})$). 
    
    \item The Chern number 
    \begin{equation}
        {\cal C}=-\frac{1}{2\pi}\oint_{{\cal S}}\Omega\cdot\hat{\bf n}dS
    \end{equation}
    measures the flux of the Berry curvature through the closed surface ${\cal S}$ in the image space of parametrized by ${\bf d}({\bf k})$, whose domain is given by the BZ torus. 
    
    \item The three possible values for the Chern number are 
    \begin{equation}
        {\cal C}=1,0,-1,
    \end{equation}
    depending on whether or not the closed surface formed by the mapping of the BZ torus through the vector field ${\bf d}({\bf k})$ contains the origin. 

    \item Topological transitions can be induced by variation of the band-splitting parameter $M$.
\end{itemize}

\section{S3: Evolution of band parameters}

A perfect crystal is characterized by very intense and sharp peaks in the Fraunhofer diffraction pattern of Bragg scattering. The intensity is proportional to the lattice structure factor, $S({\bf k}^\prime-{\bf k})=1/N\sum_{{\bf r}_i,{\bf r}_j} f_{{\bf r}_i} f_{{\bf r}_j} e^{i({\bf k}^\prime-{\bf k})\cdot({\bf r}_i-{\bf r}_j)}$, where $f_{{\bf r}_i}$ are atomic form factors. For a pristine crystal $f_{{\bf r}_i}=f_{{\bf r}_j}=1$ and thus $S({\bf k}^\prime-{\bf k})=\sum_{{\bf g}}\delta_{{\bf k}^\prime-{\bf k},{\bf g}}$, where ${\bf g}$ is a reciprocal lattice vector. The Fraunhoffer diffraction pattern in this case corresponds to $\delta-$like peaks at ${\bf k}^\prime={\bf k}+{\bf g}$. In the random atom gas limit, however, $f_{{\bf r}_i} f_{{\bf r}_j}=\delta_{{\bf r}_i,{\bf r}_j}$, and $S({\bf q})=1$, and the Fraunhoffer diffraction pattern in this case corresponds to an isotropic disc of even intensity.

For a particular electron of wave number ${\bf k}$ and orbital character $\alpha$, pseudopotential perturbation theory gives us \cite{harrison1989electronic}
\begin{equation}
    E^{\alpha}({\bf k})=E^{\alpha}_0({\bf k})+w^{\alpha}_0+
    \sum_{{\bf k}^\prime}\frac{S^{\alpha\alpha}({\bf k}^\prime-{\bf k})|w_{{\bf k}^\prime-{\bf k}}|^2}{E^{\alpha}_0({\bf k})-E^{\alpha}_0({\bf k}^\prime)},
\end{equation}
where $\langle{\bf k}^\prime|W^{\alpha}({\bf r}_j)|{\bf k}\rangle=(1/\sqrt{N})f^{\alpha}_{{\bf r}_j}e^{i({\bf k}^\prime-{\bf k})\cdot{\bf r}_j}w_{{\bf k}^\prime-{\bf k}}$, and $f^{\alpha}_{{\bf r}_j}$ is a form factor for the orbital $\alpha=s,p$ at the atomic site ${\bf r}_j$. In the above expression we also have $S^{\alpha\alpha}({\bf k}^\prime-{\bf k})=(1/N)\sum_{{\bf r}_i,{\bf r}_j}f^{\alpha}_{{\bf r}_i}f^{\alpha}_{{\bf r}_j}e^{i({\bf k}^\prime-{\bf k})\cdot({\bf r}_i-{\bf r}_i)}$ as the structure factor and $w_{{\bf k}^\prime-{\bf k}}$ is the Fourier transform of the lattice potential. The band-splitting, $M$, corresponds to the difference in the onsite energy between the $s-$like and $p-$like bands, $M=\varepsilon^s-\varepsilon^p$. In order to calculate the onsite energies, $\varepsilon^\alpha$ for $\alpha=s,p$, we shall make use the free electron energy at the $\Gamma$ point and write $E^\alpha_0({\bf k})=\hbar^2{\bf k}^2/2m^*_\alpha$, where $m^*_\alpha$ is the effective mass at the $\alpha=s,p$ band. The pseudopotential contribution to the onsite band energy, $\varepsilon^\alpha(\sigma)$, can be obtained by summing up the last term of the above expression, multiplying by $2$ to account for spin degeneracy and dividing by $N$, or
\begin{equation}
    \varepsilon^\alpha(\sigma)=\varepsilon_0^\alpha-\frac{2}{N}\frac{2 m^*_i}{\hbar^2}\frac{V}{8\pi}
    \sum_{{\bf q}\neq {\bf g}}S^{\alpha\alpha}_\sigma({\bf q})|w_{{\bf q}}|^2F(q),
\end{equation}
where $V=Na$ is the volume of the crystal in terms of the lattice constant, $a$, ${\bf q}={\bf k}^\prime-{\bf k}$ and the function $F(q)=1-\Theta(q-2g)\sqrt{1-(2g/q)^2}$ was obtained after integration over ${\bf k}$ for the entire BZ. Here $\varepsilon_0^i$ represents the onsite energy for the pristine case ${\bf q}\equiv{\bf g}$, $S^{\alpha\alpha}_\sigma({\bf q})$ stands for Hosemann disordered structure factor and $g$ is the size of the primitive vector of the reciprocal lattice. The dependence of the band-splitting on the disordering of the crystal is then governed by the equation

%
%\begin{eqnarray}
%    \sum_{{\bf k}}\frac{1}{{\bf k}^2-({\bf k}+{\bf q})^2}&\rightarrow&-
%    V\int\frac{d^2{\bf k}}{(2\pi)^2}\frac{1}{{\bf q}^2+2{\bf k}\cdot{\bf q}}\nonumber\\
%    &=&-\frac{V}{4\pi^2}\int_0^g\, dk\; k\int_0^{2\pi}\frac{d\theta}{q^2+2kq \cos{\theta}}\nonumber\\
%    &=&-\frac{V}{4\pi^2 q^2}\int_0^g\, dk\; k\int_{-1}^{+1}\frac{d\nu}{\sqrt{1-\nu^2}}\frac{1}{1+\frac{2k\nu}{q}}\nonumber\\
%    &=&-\frac{V}{4\pi^2 q^2}\int_0^g\, dk\; k\frac{\pi q}{\sqrt{q^2-4k^2}},\quad\mbox{if $q\geq 2k$}\nonumber\\
%    &=&-\frac{V}{4\pi q}\int_0^g\, dk\frac{k}{\sqrt{q^2-4k^2}}\nonumber\\
%    &=&-\frac{V}{8\pi}\left[1-\sqrt{1-(2g/q)^2}\right],
%\end{eqnarray}
% 

%
\begin{equation}
    M(\sigma)=M(0)-\frac{(m^*_s-m^*_p)a_0^2}{2\pi\hbar^2}
    \sum_{{\bf q}\neq {\bf g}}S^{\alpha\alpha}_\sigma({\bf q})|w_{{\bf q}}|^2F(q),
\end{equation}
and we see that for $m^*_s-m^*_p>0$ the band-splitting $\Delta M(\sigma)$ is a monotonic decreasing function of the disorder variance $\sigma$, that can cross the $\Delta M(\sigma_c)=0$ value where the band inversion takes place and the topological gap closes. Next we introduce a sharp cutoff that splits the structure factor into two pieces: pristine, where $S_<({\bf q})=\sum_{{\bf g}}\delta_{{\bf q},{\bf g}}$ for $g<g_c$, and amorphous where $S_>({\bf q})=1$, for $g>g_c$. Finally, we shall use for the reciprocal lattice potential the expression $w_{{\bf q}}=1/(q^2+\kappa^2)$. If we recall that for $q<2g$ the function $F(q)=1$, the equation for the onsite energy then becomes
\begin{eqnarray}
    M(\sigma)&=&M(0)-\frac{(m^*_s-m^*_p)a_0^2}{2\pi\hbar^2}
    \int_0^{\delta g}\frac{d^2{\bf q}}{(2\pi)^2}|w_{{\bf q}}|^2 F(q)\nonumber\\
    &=&M(0)-\frac{(m^*_s-m^*_p)a_0^2}{4\pi^2\hbar^2}
    \int_0^{\delta g}dq\frac{q}{q^2+\kappa^2}\nonumber\\
    &=&M(0)+\frac{(m^*_s-m^*_p)a_0^2}{8\pi^2\hbar^2}\log{\left[\frac{\kappa^2}{\kappa^2+\delta g^2}\right]}\nonumber\\
    &\rightarrow&M(0)-\frac{(m^*_s-m^*_p)a_0^2}{8\pi^2\hbar^2}\frac{\delta g^2}{\kappa^2}.
\end{eqnarray}

\section{S4: Nature of the topological states}

\begin{figure}
\centering
\includegraphics[width=0.99\columnwidth]{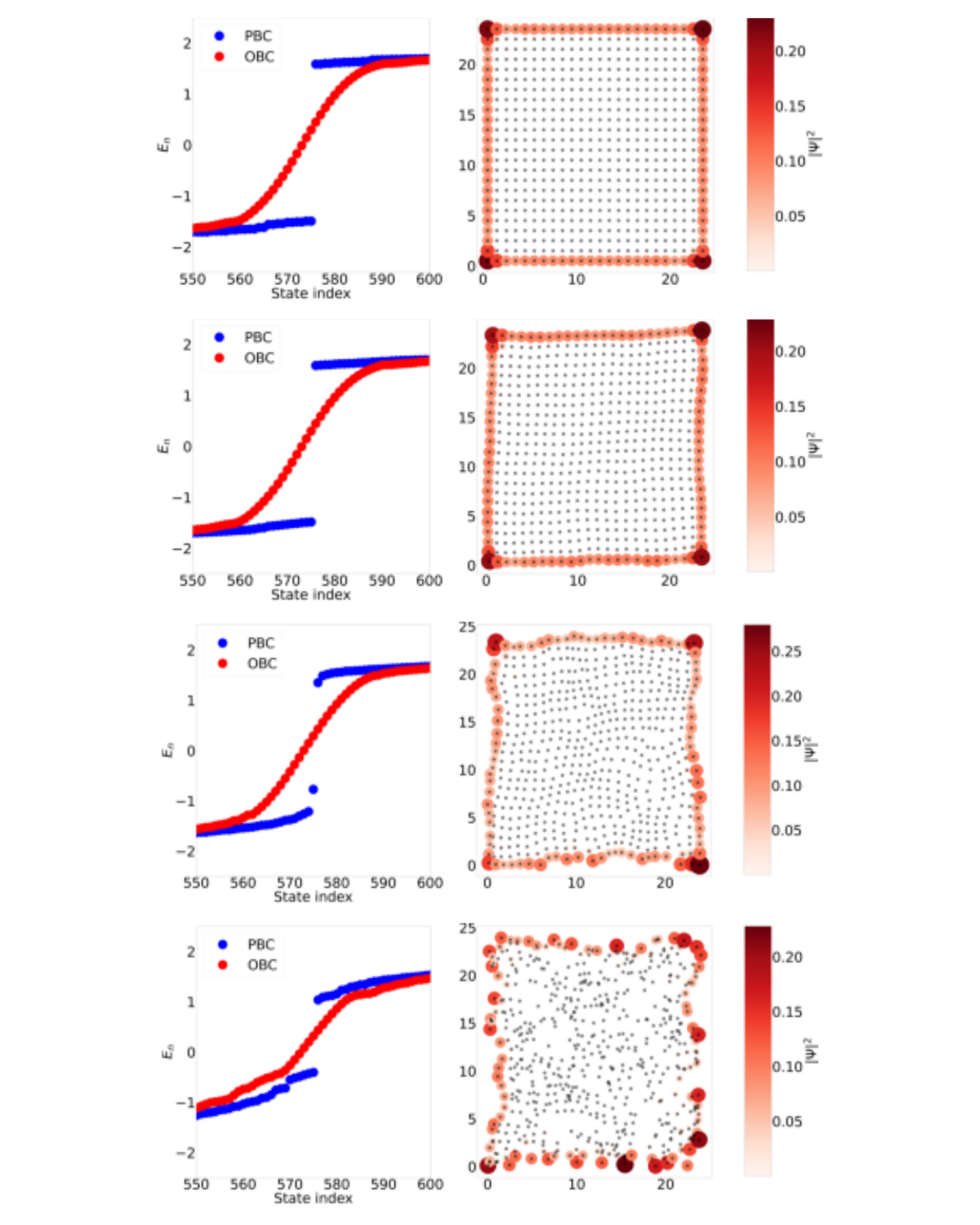}
\caption{Energy eigenvalues and local density of states (LDOS) for four different realizations of the structures through the PN procedure. From top to bottom increasing structural disorder. The left column shows the midgap states in the open boundary conditions (OBC), in contrast to the insulating gap in periodic boundary condition (PBC). The right column is the LDOS calculated for the midgap states in each case. The color code shows the probability $|\psi|^2$ for finding an electron in each site. The localization on the borders is evident in all cases. The other parameters are kept the same as in the main text: $t_2 = 0.25$ and $R = 4.0$.}
\label{LDOS}
\end{figure}

In the main text, we introduce the model Hamiltonian used to describe the dynamics of the topological states during the structural changes of the lattice. For characterization of such states, we used the Bott index, which shows non-trivial values through a wide range of parameters, as shown in Fig.~3 (in the main text). In order to further investigate the states arising from the model and to ensure their topological nature, we calculate the energy eigenvalues for a given realization of the lattices through the amorphization procedure, in both the periodic boundary conditions (PBC) and open boundary conditions (OBC) cases. In the left panel of Fig. \ref{LDOS} we show the results of these calculations. It is clear that when the system has PBC, a gap opens and the system is in a insulating state. However, when OBC are imposed, the gap closes and we observe the presence of midgap states. In the right panel of Fig. \ref{LDOS}, we show the usual nature of these midgap states by calculating the local density of states (LDOS) in the sites of the lattices. We can observe that these midgap states are edge states, localized on the borders of the lattice. This characteristic is not altered by the amorphization, the edge states in the crystalline phase are robust against the increasing disorder (decreasing correlation lenght) and keep their nature even in the completely random lattice. The parameters used for this calculation are set to $M = -0.5$ and $\lambda = 0.5$, which locate these states in a non-trivial region of the phase diagram along the amorphization, with $\mathcal{B} = +1$. This value of the local topological marker, together with the edge nature of these midgap states, ensure the topological nature of the states under study.

%%%%%%%%%%%%%%%%%%%%%%%%%%%%%%%%%%%%%%%%%%%%%%%%%%%%%%%%%
\bibliography{amorphousTI}
%%%%%%%%%%%%%%%%%%%%%%%%%%%%%%%%%%%%%%%%%%%%%%%%%%%%%%%%%